\documentclass[fleqn,usenatbib]{mnras}

\usepackage[T1]{fontenc}

\DeclareRobustCommand{\VAN}[3]{#2}
\let\VANthebibliography\thebibliography
\def\thebibliography{\DeclareRobustCommand{\VAN}[3]{##3}\VANthebibliography}

\usepackage{graphicx}	
\usepackage{amsmath}	
\usepackage{amssymb}	
\usepackage{newtxtext,newtxmath}
\usepackage{pdflscape}
\usepackage{subcaption}
\usepackage{xcolor}

\DeclareMathOperator{\arctantwo}{arctan2}

\newcommand{\msun}{{\rm ~M}_\odot}	
\newcommand{\kpc}{{\rm ~kpc}}	    
\newcommand{\kms}{{\rm \,km\,s^{-1}}}

\title[The OC stream in deforming dark matter haloes]{The effect of the deforming dark matter haloes of the Milky Way and the Large Magellanic Cloud on the Orphan-Chenab stream}

\author[S. Lilleengen et al.]{
\parbox{\textwidth}{
\Large
Sophia~Lilleengen,$^{1}$\thanks{\href{mailto:s.lilleengen@surrey.ac.uk}{s.lilleengen@surrey.ac.uk}}
Michael~S.~Petersen,$^{2,3}$
Denis~Erkal,$^{1}$
Jorge~Pe\~narrubia,$^{3}$
Sergey~E.~Koposov,$^{3,4,5}$
Ting~S.~Li,$^{6}$
Lara~R.~Cullinane,$^{7}$
Alexander~P.~Ji,$^{8,9}$
Kyler~Kuehn,$^{10,11}$
Geraint~F.~Lewis,$^{12}$
Dougal~Mackey,$^{13}$
Andrew~B.~Pace,$^{14}$
Nora~Shipp,$^{15}$
Daniel~B.~Zucker,$^{16,17}$
Joss~Bland-Hawthorn,$^{12,18}$
and Tariq~Hilmi$^{1}$
\begin{center} (\(S^5\) Collaboration) \end{center}
}
\vspace{0.4cm}
\\
\parbox{\textwidth}{
$^{1}$ Department of Physics, University of Surrey, Guildford GU2 7XH, UK\\
$^{2}$ CNRS and Sorbonne Universite, UMR 7095, Institut d’Astrophysique de Paris, 98 bis Boulevard Arago, F-75014 Paris, France\\
$^{3}$ Institute for Astronomy, University of Edinburgh, Royal Observatory, Blackford Hill, Edinburgh EH9 3HJ, UK\\
$^{4}$ Institute of Astronomy, University of Cambridge, Madingley Road, Cambridge CB3 0HA, UK\\
$^{5}$ Kavli Institute for Cosmology, University of Cambridge, Madingley Road, Cambridge CB3 0HA, UK\\
$^{6}$ Department of Astronomy and Astrophysics, University of Toronto, 50 St. George Street, Toronto ON, M5S 3H4, Canada\\
$^{7}$ Department of Physics and Astronomy, Johns Hopkins University, 3400 N. Charles St, Baltimore, MD 21218, USA\\
$^{8}$ Department of Astronomy \& Astrophysics, University of Chicago, 5640 S Ellis Avenue, Chicago, IL 60637, USA\\
$^{9}$ Kavli Institute for Cosmological Physics, University of Chicago, Chicago, IL 60637, USA\\
$^{10}$ Lowell Observatory, 1400 W Mars Hill Rd, Flagstaff,  AZ 86001, USA\\
$^{11}$ Australian Astronomical Optics, Faculty of Science and Engineering, Macquarie University, Macquarie Park, NSW 2113, Australia\\
$^{12}$ Sydney Institute for Astronomy, School of Physics, A28, The University of Sydney, NSW 2006, Australia\\
$^{13}$ Research School of Astronomy and Astrophysics, Australian National University, Canberra, ACT 2611, Australia\\
$^{14}$ McWilliams Center for Cosmology, Carnegie Mellon University, 5000 Forbes Ave, Pittsburgh, PA 15213, USA\\
$^{15}$ MIT Kavli Institute for Astrophysics and Space Research, 77 Massachusetts Ave., Cambridge, MA 02139, USA\\
$^{16}$ School of Mathematical and Physical Sciences, Macquarie University, Sydney, NSW 2109, Australia\\
$^{17}$ Macquarie University Research Centre for Astronomy, Astrophysics \& Astrophotonics, Sydney, NSW 2109, Australia\\
$^{18}$ Centre of Excellence for All-Sky Astrophysics in Three Dimensions (ASTRO 3D), Australia\\
}
}

\date{Accepted XXX. Received YYY; in original form ZZZ}

\pubyear{2022}

\begin{document}
\label{firstpage}
\pagerange{\pageref{firstpage}--\pageref{lastpage}}
\maketitle

\begin{abstract}
It has recently been shown that the Large Magellanic Cloud (LMC) has a substantial effect on the Milky Way's stellar halo and stellar streams. Here, we explore how deformations of the Milky Way and LMC's dark matter haloes affect stellar streams, and whether these effects are observable. In particular, we focus on the Orphan-Chenab (OC) stream which passes particularly close to the LMC and spans a large portion of the Milky Way's halo. We represent the Milky Way--LMC system using basis function expansions that capture their evolution in an $N$-body simulation. We present the properties of this system, such as the evolution of the densities and force fields of each galaxy. The OC stream is evolved in this time-dependent, deforming potential, and we investigate the effects of the various moments of the Milky Way and the LMC. We find that the simulated OC stream is strongly influenced by the deformations of both the Milky Way and the LMC and that this effect is much larger than current observational errors. In particular, the Milky Way dipole has the biggest impact on the stream, followed by the evolution of the LMC's monopole, and the LMC's quadrupole. Detecting these effects would confirm a key prediction of collisionless, cold dark matter, and would be a powerful test of alternative dark matter and alternative gravity models. 
\end{abstract}

\begin{keywords}
Galaxy: evolution -- Galaxy: halo -- Galaxy: kinematics and dynamics  -- Galaxy: structure -- Magellanic Clouds
\end{keywords}


\section{Introduction}\label{sec:intro}
Despite its ubiquity, dark matter continues to evade direct \citep[e.g.][]{Aprile+2018}, indirect \citep[e.g.][]{Gaskins2016}, and collider searches \citep[e.g.][]{Kahlhoefer2017}. To date, the only evidence of dark matter has come from its gravitational effect on astrophysical probes \citep[e.g.][]{Zwicky1937,Rubin+1970,Read2014,Planck+2016}. Merging galaxy clusters, which effectively serve as huge dark matter colliders, have proven a particularly fruitful testing ground for the collisionless nature of dark matter \citep[e.g.][]{Clowe+2004,Markevitch+2004,Bradac+2008,Jee+2014}. The most iconic of these is the Bullet Cluster \citep[][]{Markevitch+2002} which shows a clear offset between the dark matter (as measured with weak lensing) and the gas (as measured with X-rays) which has been used both as evidence of dark matter \citep[e.g.][]{Clowe+2004} and to set constraints on self-interacting dark matter \citep[e.g.][]{Markevitch+2004,Robertson+2017}. 

Evidence is mounting that the ongoing merger of the Large Magellanic Cloud (LMC) and the Milky Way may serve as a similarly useful dark matter collider. The LMC is believed to be on its first approach to the Milky Way \citep[][]{Besla+2007} and appears to still have a substantial dark matter halo consistent with what is expected from abundance matching, $\sim 2\times10^{11} \msun$ \citep[e.g.][]{Moster+2013,Behroozi+2013}. Such a massive halo would have a large effect on structures in the Milky Way, and such effects have recently been detected. For example, in order to explain the nearby presence of the Small Magellanic Cloud (SMC) and other Magellanic satellites, an LMC mass of $>10^{11} \msun$ is needed \citep[e.g.][]{Kallivayalil+2013,Kallivayalil+2018,Erkal+2020,Patel+2020}. An LMC mass of $\sim 2.5\times10^{11} \msun$ is needed to explain the timing argument with M31 and the nearby Hubble flow \citep{Penarrubia+2016}. The LMC's effect has also been detected in the Milky Way's stellar halo, both in terms of kinematics \citep{Erkal...MWbias2...21,Petersen...reflexmotion...21} and overdensities in the stellar halo \citep[e.g.][]{Belokurov...Pisces...19,GaravitoCamargo...DMwake...19,Conroy...21}, all consistent with LMC masses of $\sim (1-2)\times10^{11} \msun$. Finally, the LMC has perturbed many stellar streams in the Milky Way, allowing for a precise measurement of its mass, $\sim(1.3-1.9)\times10^{11}\msun$ \citep[][]{Erkal..Orphan..2019,koposov19,Shipp...DESpms...19,Shipp...LMCmassS5...2021,Vasiliev...tango3....2021}.

These large LMC masses correspond to roughly $10-20$~per~cent of the Milky Way mass \citep[e.g.][]{Wang+2020}. As such, this substantial merger will create significant tidal deformations in the dark matter haloes of both the Milky Way and LMC \citep[e.g.][]{Weinberg1989,Weinberg1998,Laporte+2018,GaravitoCamargo...DMwake...19,GaravitoCamargo...BFE...21}. \cite{GaravitoCamargo...BFE...21} quantified these deformations by first running an $N$-body simulation of the Milky Way--LMC encounter and then fitting basis function expansions (BFE) to the present-day snapshot of a Milky Way--LMC realisation. Their analysis showed that the dark matter haloes of both the Milky Way and the LMC deform substantially. For their Milky Way model, these effects were comparable to the expected halo triaxiality seen in cosmological simulations \citep[e.g.][]{Chua+2019} and thus would need to be understood in order to robustly measure the shape of the Milky Way halo. Furthermore, previous works have shown that the detailed structure of these deformations would also depend on the nature of dark matter \citep[e.g.][]{Furlanetto+2002,hui+2017,lancaster+2020}.  

In this work, we show that stellar streams are sensitive to the deforming dark matter haloes of both the Milky Way and the LMC. Stellar streams form as globular clusters and dwarf galaxies disrupt in the presence of their host galaxy. They are powerful probes of the host's gravitational potential \citep{Johnston...streams...1999,Helmi+1999}: the collection of stars in the stream roughly delineate orbits in the host potential \citep{Sanders+2013}, allowing us to infer the accelerations that the stars experience (and hence the host's gravitational field) without having to directly measure the acceleration\footnote{See, however, \cite{Quercellini+2008,Silverwood+2019,Chakrabarti+2020} for efforts to directly measure accelerations of bright, nearby stars with upcoming spectrographs, \cite{Klioner+2021} for a measurement of the solar system's acceleration with quasars, and \cite{Chakrabarti+2021} for a measurement of the local acceleration within $\sim 1$ kpc with binary pulsars.}. Many streams in the Milky Way have already been used to fit the Galactic potential  \citep[e.g.][]{ Law...Majewski...2010, Koposov...GD1...2010, Vera-Ciro...SagLMC...13, Bonaca...streams..2014, Gibbons...mLCS...2014, Kuepper...Pal5...2015, Bovy...streams...2016, Erkal..Orphan..2019, Malhan...GD1...2019, Vasiliev...tango3....2021}. 

In this study, we focus on the Orphan-Chenab (OC) stream \citep{Grillmair...Orphan...2006, Belokurov...streams...2006,Shipp...DES...18,koposov19}. This stream is particularly well-suited to study the deformations of the Milky Way and LMC since it has experienced a close passage with the LMC \citep[$\sim$15 kpc,][]{Erkal..Orphan..2019} and spans a large portion of the Milky Way \citep{koposov19}. \cite{Erkal..Orphan..2019} have used this stream to measure the potential of both the Milky Way and the LMC. Their fits prefer a Milky Way dark matter halo that is misaligned with the Milky Way's disc and either strongly oblate or strongly prolate. Interestingly, the oblate halo is roughly aligned with the orbital plane of the LMC and the prolate halo is aligned with the present-day position of the LMC, hinting that the inferred halo shape may be connected to the LMC. Furthermore, the oblate halo of \cite{Erkal..Orphan..2019} is similar to the triaxial (but nearly oblate) haloes inferred with the Sagittarius stream \citep[][]{Law...Majewski...2010,Vasiliev...tango3....2021}. We note that while the fits in \cite{Erkal..Orphan..2019} and \cite{Vasiliev...tango3....2021} allow the Milky Way to move in response to the LMC, both of these fits assume that the dark matter haloes of the Milky Way and LMC are rigid (i.e. time-independent)\footnote{We note that \citet{Vasiliev...tango3....2021} performed the MCMC exploration of the Milky Way and LMC parameters using rigid haloes. They then ran live $N$-body simulations and provided time-dependent potentials for some sets of the best-fit parameters.}. This leads to the question; how exactly are streams affected by the deformation of the dark matter haloes of the Milky Way and the LMC? 

In order to understand how the OC stream is affected by these deformations, we study the interaction of the Milky Way and the LMC with $N$-body simulations. 
The $N$-body simulations use basis function expansions to compute the density, potential, and forces that each particle experiences as a function of time and are performed using the basis function expansion software toolkit \textsc{exp} \citep{petersen..exp..21}. 
This provides us with time-dependent BFEs that we can use to evolve stellar streams in this disturbed system. 
The time-dependent BFEs allow us to explore how the different orders of both the Milky Way and LMC expansion (in particular the multipole orders) affect the stream.

This paper is organised as follows. 
In Section~\ref{sec:BFEs}, we describe the BFE technique and our Milky Way--LMC model. 
In Section~\ref{sec:streams}, we present the OC stream data and models and describe how they are affected by different moments of the deforming Milky Way--LMC model.
Then in Section~\ref{sec:discussion}, we discuss our results, how tracing these deformations with stellar streams could allow us to distinguish between several dark matter and alternative gravity models, and the possible influence of the SMC on the Milky Way--LMC model. We summarise our findings and conclude in Section~\ref{sec:conclusions}.

\section{Basis Function Expansions}\label{sec:BFEs}

\subsection{\textsc{exp}}\label{sec:BFEs_EXP}

To model orbits for stellar stream particles in the combined (and evolving) Milky Way--LMC environment, we require a description of the potential and forces at any arbitrary point in the system (and through time). 
Unfortunately, analytic potentials will not describe deformations in the dark matter haloes of the Milky Way or LMC. 
We, therefore, seek a flexible alternate method to describe the density, potential, and forces as they evolve through time: basis function expansions. Basis function expansions have proved a viable method to produce flexible models of the Milky Way \citep[e.g.][]{Petersen...shadowbar...2016,Dai...BFE...2018,Petersen...bars...2019,Petersen...dipole...2020,GaravitoCamargo...BFE...21}.
We use the basis function expansion machinery implemented in {\sc exp} \citep{petersen..exp..21} to both run $N$-body simulations, as well as to resimulate the fields ex post facto.

Briefly, basis function expansions model a target distribution as the sum of {\it orthogonal basis functions}, represented by the index $\mu$, each of which adds an additional degree of freedom to the system. 
Each function has an associated {\it coefficient}, $A_\mu$, which is the contribution of the function to the total description of the system.
The total system at any given time is parameterised by the functions and their coefficients.
We vary the coefficients through time to describe the evolving potential and keep the functions fixed in their initial forms.

In the case of a three-dimensional distribution of masses, one forms biorthogonal potential-density pairs that solve Poisson's equation\footnote{That is, pairs of indexed potential and density functions  $(\phi_i(\mathbf{x}),\rho_i(\mathbf{x}))$  satisfy the following equations: $\nabla^2 \phi_i = 4\pi G \rho_i$ and $\int d\mathbf{x}\, \phi_i(\mathbf{x}) \rho_j(\mathbf{x}) = 4\pi G\delta_{ij}$, where $\delta_{ij}$ is the Kronecker delta.}. 
The Hernquist basis set \citep{hernquist92} is one such set of basis functions, which efficiently expands the Hernquist density distribution \citep{hernquist90}. 
Another standard method to expand an input model spherical distribution, $\rho_{\rm model}(r,\phi,\theta)$, uses spherical harmonics $Y_{l}^{m}$ to describe the $(\phi,\theta)$ angular dependence, and eigenfunctions of the Sturm-Liouville equation \citep[of which Poisson's equation is a special case, see][]{weinberg99} to describe the three-dimensional radial dependence (indexed as $n$).
The radial basis index, $n$, corresponds to the number of nodes in the radial function. For $l=0$, $n$ is equal to the number of nodes in each radial function. For $l>0$, the number of nodes in the radial function is $n+1$.
Each spherical basis function may be represented by the triple $\mu\equiv(l,m,n)$, and the entire spherical coefficient set is of size $(l_{\rm max}+1)^2\cdot (n_{\rm max}+1)$ at each timestep.
Previous work using {\sc exp} demonstrated high force reconstruction accuracy when modelling the Milky Way using $l_{\rm max}=6,\,n_{\rm max}=17$ for the halo \citep{petersen..exp..21}. As the LMC is undergoing a more significant deformation than the Milky Way, we model the LMC using a modestly larger number of radial functions to resolve the deformation. Guided by previous models for the Milky Way and LMC \citep{Petersen...reflexmotion...21}, we choose $l_{\rm max}=6,\,n_{\rm max}=17$ for the Milky Way and $l_{\rm max}=6,\,n_{\rm max}=23$ for the LMC\footnote{For all expansions, we index the lowest-order radial function as $n=0$, meaning that the Milky Way (LMC) has 18 (24) radial functions per harmonic order. The harmonic functions (indexed by $l$ in the spherical case and $m$ in the cylindrical case) both also begin at 0.}. 
The basis functions are defined to be biorthogonal \citep{weinberg99}, such that the inner product of the density and potential functions for each component is the sum of the power computed from the entire coefficient set. 
The physical interpretation of the coefficient power is then that of gravitational energy, which allows one to interpret power as the self-gravity of the system represented by the different functions. 
Throughout, we will describe individual {\it harmonic subspaces}, referring to the $l=0$ terms as the monopole, the $l=1$ terms as the dipole, and the $l=2$ terms as the quadrupole\footnote{Higher $l$ orders will be simply described as $l=3,\dots,l_{\rm max}$.}. 
The lowest-order monopole function ($l=n=0$) is tailored to match the density profile of the target mass distribution, such that $\rho_{lmn}(r)=\rho_{000}(r)=\rho_{\rm model}(r)$ (and equivalent relationships for the potential $\Phi$).

While a spherical object, such as the dark matter haloes of the Milky Way and LMC, may be straightforwardly expanded using a spherical basis, a cylindrical mass distribution such as a stellar disc would necessitate extremely high $l$ orders to approximate a thin structure. 
We, therefore, use the adaptive basis technique in {\sc exp} to derive a basis that more closely matches a target stellar density distribution. 
The cylindrical basis is selected from a high-order ($l_{\rm max}=64, n_{\rm max}=63$) spherical expansion by finding the optimal meridional functions through eigendecomposition. 
The cylindrical basis functions are  described by two cylindrical coordinates, $R$ and $z$, with the angular dependence coming from a Fourier expansion in azimuthal harmonics $m$. 
As in the case of the spherical expansions, the lowest-order monopole function ($m=n=0$) will closely resemble the equilibrium density of the target mass distribution, $\rho_{mn}(R,z,\phi)=\rho_{00}(R,z,\phi)\approx\rho_{\rm model}(R,z)$.
For the total cylindrical expansion, we retain functions and coefficients up to $m_{\rm max}=6$ and $n_{\rm max}=17$.
The cylindrical coefficient set is then of size $(2m_{\rm max}+1)\cdot(n_{\rm max}+1)$ at each timestep.

Representing the system as the linear sum of solutions to Poisson's equation is a so-called {\it global basis}, that has the principal benefits of describing the self-gravity in a given correlated evolutionary mode of the system. 
The downside of global bases is their susceptibility to aliasing\footnote{Aliasing formally owes to the truncation of the infinite series; in practice, regions with low sampling (i.e. low density) are the most affected. 
Fortunately, these regions are low-density and therefore tend to have little impact on the evolution, but may cause structures that are obvious by eye. In an idealised test of a deformed model, \protect{\cite{petersen..exp..21}} demonstrate that regions at $r>5r_s$, mismatch in the outskirts of models may contribute a 1\% error to the forces. These regions do not strongly affect the evolution of the OC stream.}. 
Further, owing primarily to finite-$N$ effects in the models, there is uncertainty on the coefficients of a given basis function. In practice, this will add numerical noise to the (re-)simulation, as well as create scatter in the coefficients.

Despite these limitations, careful use of basis function expansions provides a computationally efficient and flexible means to describe an evolving model. 
In the next Section, we introduce the models from which we obtain the basis functions and coefficients in order to model the OC stream.

\subsection{\textit{N}-body models}

We build our model Milky Way--LMC from three components: a Milky Way stellar component (a disc and bulge), a Milky Way dark matter halo, and an LMC dark matter halo.
The dark matter profiles are selected to be the best fit, but reflexive, spherical potential (labelled `sph. rMW+LMC') from table A1 of \citet{Erkal..Orphan..2019}: an NFW \citep{navarro96} Milky Way dark matter halo with $M_{\rm MW~halo}=7.92\times10^{11}\msun$, $r_s=12.8\kpc$, and $c=15.3$, a Miyamoto-Nagai \citep{miyamoto75} stellar disc with $M_{\rm MW~disc}=6.8\times10^{10}\msun$, $a=3.0 \kpc$, and $b=0.28 \kpc$, a Hernquist \citep{hernquist90} stellar bulge with $M_{\rm MW~bulge}=5\times10^{9}\msun$, $r_s=0.5 \kpc$, and a Hernquist LMC dark matter halo with $M_{\rm LMC}=1.25\times10^{11}\msun$ and $r_s=14.9~\kpc$. 
While we do not apply a truncation to the Hernquist profile of the LMC, we add a truncation to the Milky Way halo potential to counteract the infinite mass of the NFW potential. 
We truncate the profile by multiplying the density profile by a normalised error function such that the final Milky Way profile is $\rho_{\rm Milky~Way}(r)=0.5\rho_{\rm NFW}(r)\left(1-{\rm erf}\left[(r-r_{\rm trunc})/w_{\rm trunc}\right]\right)$ where $r_{\rm trunc}=430\kpc$ and $w_{\rm trunc}=54\kpc$.
The models are realised with $N_{\rm MW~halo}=10^7$, $N_{\rm LMC}=10^7$, $N_{\rm MW~disc}=10^6$ particles, following the procedures in \citet{petersen..ics..2021}. 
We define the bases for each component using the basis selection techniques in {\sc exp}. The Milky Way and LMC dark matter haloes are represented as spherical bases; the Milky Way stellar disc and bulge mass distributions are combined and represented by a single cylindrical basis, as introduced in Section~\ref{sec:BFEs_EXP}.

The $N$-body models are evolved using {\sc exp}, which uses the basis functions for each component to obtain the forces \citep[see][for details]{petersen..exp..21}. 
We have checked that the distance between the expansion centre of the Milky Way halo and the Milky Way stellar component (consisting of the disc and the bulge) is smaller than the minimum node spacing in the Milky Way halo expansion (i.e. the node spacing in the $l=0$, $n=17$ Milky Way basis function).
To realise the trajectory, we first run point-mass models of the Milky Way and LMC in reverse from their present-day locations to obtain a rough trajectory. We specify the present-day centre position and velocity of the LMC using \(\left(\alpha_{\rm LMC},\delta_{\rm LMC}\right)=\left(78.76^\circ\pm0.52,-69.19^\circ\pm0.25\right)\), with proper motions \(\left(\mu_{\alpha^\star,{\rm LMC}}, \mu_{\delta,{\rm LMC}}\right)=\left(-1.91\pm0.02~{\rm mas/yr}, 0.229\pm0.047~{\rm mas/yr}\right)\) from \citet{Kallivayalil+2013}, the distance from \citet{Pietrzynski_2019}, \(d_{\rm LMC} = 49.59\pm0.54~{\rm kpc}\), and the line-of-sight velocity from \citet{vanderMarel_2002}, \(v_{\rm los,~LMC}=262.2\pm3.4\kms \).
We find that we recover a qualitatively similar orbit to the best-fit reflexive spherical model from \citet{Erkal..Orphan..2019}. 
Individual model realisations may be run on modest-sized supercomputers owing to the computational efficiency of {\sc exp}.
However, in order to obtain a realisation that is sufficiently close to the present-day Milky Way--LMC pair, we run a grid of 15 live $N$-body models around the rewound point mass models and select the model which best matches the present-day observables. 
We note that these live $N$-body models self-consistently include the dynamical friction the LMC experiences in the presence of the Milky Way, in contrast to the initial point-mass models, which do not include any dynamical friction prescription.
Future models may reach quantitatively better matches to the luminous positions of the Milky Way and LMC, but for our purposes, the match of the Milky Way--LMC pair is sufficient to study the effect of deformations on stellar stream observables.

For computational convenience, we convert the physical units of the simulation into virial units ($G=T_{\rm virial}=M_{\rm MW}=1$). 
In these units, we allow maximum timesteps of ${\rm d}t_{\rm max}=0.002~T_{\rm virial}$, with smaller timesteps decided by adaptive criteria, down to a minimum of ${\rm d}t_{\rm min}=0.000125~T_{\rm virial}$. 
At each minimum timestep in the simulation, the coefficients for each function are tabulated from the present distribution of the particles and recorded for later use.
The live simulation starts at $T=-1.2~T_{\rm virial}(=-2.5~{\rm Gyr})$ before the present day ($T=0$). 
When representing the density/force/potential fields of the system at times prior to the start of the live simulation, we set the coefficients for each component to be their initial values, and place the LMC on a coasting orbit determined by extrapolating the $T=-1.2~T_{\rm virial}$ velocity vector of the LMC backwards in time \footnote{A {\sc python} interface to integrate orbits and access the expansion model for the simulation can be found here: \href{https://github.com/sophialilleengen/mwlmc}{https://github.com/sophialilleengen/mwlmc}.}. The simulation qualitatively resembles the present-day snapshots of both other simulated \citep{GaravitoCamargo...DMwake...19,GaravitoCamargo...BFE...21} and numerical \citep{Rozier+2022} models of the Milky Way--LMC interaction.

\subsection{Dipole/quadrupole evolution}\label{sec:BFE_evolution}
\begin{figure}
	\includegraphics[width=\columnwidth]{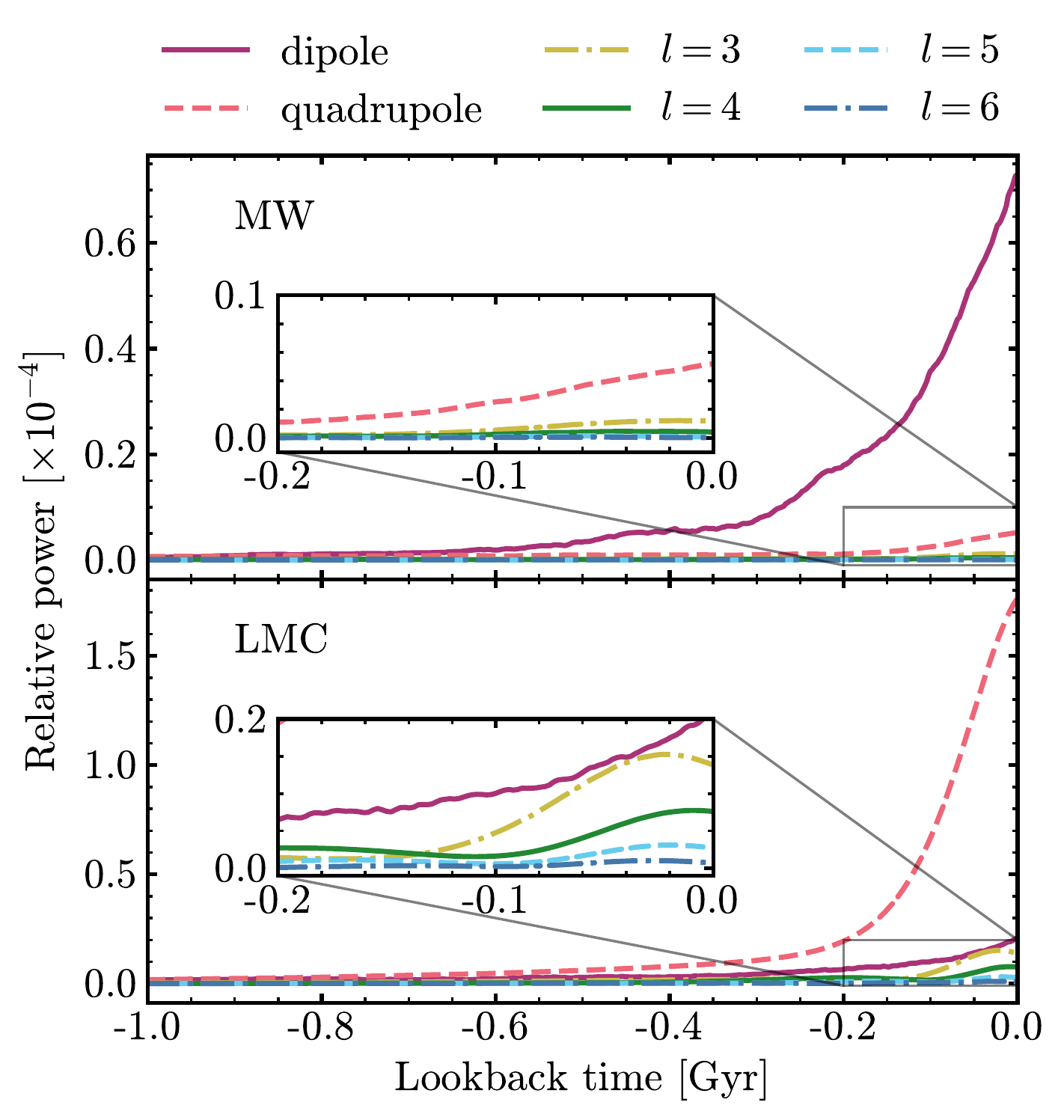}
    \caption{Time evolution of the relative power in non-monopole harmonic orders of the Milky Way halo \textit{(top)} and the LMC \textit{(bottom)}. 
    The power in each harmonic order is normalised by the monopole ($l=0$) power. 
    Power may be interpreted as the amount of self-gravity in a particular harmonic order, such that larger power values indicate more influence.
    Insets show a zoom-in focusing on the 200~Myr preceding the present day.
    The strongest multipole order in the Milky Way is its dipole which has been slowly increasing over the last 500~Myr and more rapidly over the last few hundred Myr.
    The LMC is dominated by a sharply rising quadrupole over the last 200~Myr, with contributions of the higher orders becoming nonzero in the last 100~Myr.
    }
    \label{fig:BFE_moment_evolution}
\end{figure}
Figure~\ref{fig:BFE_moment_evolution} is a global view of the Milky Way--LMC interaction, as represented by our BFE. 
To summarise the Milky Way and LMC systems, we compute the squared sum of coefficients $\sum_{n,m}|A_{lmn}|^2$ over all radial functions $n$ and $m\in[-l,\ldots,0,\ldots,l]$ for each $l$ harmonic subspace to compute a measure of the self-gravity in each harmonic subspace, $A_l$. 
The power values themselves are model-dependent, so we choose to normalise the power in each harmonic subspace to the total power in the monopole ($l=0$) harmonic subspace, $P_l/P_0 =A_{l}^2/A_{0}^2$. 
Given this normalisation, we interpret the series of power over time as the relative influence of each harmonic order to represent the global system and describe percentages of self-gravity represented by particular harmonic subspaces. 
We inspected harmonic subspaces as a function of radial order $n$ and found that the amplitudes decrease monotonically (that is, the $n=0$ order is the largest, $n=1$ the next largest, and so on). 
Therefore, we report only the sum over each harmonic subspace throughout the paper. We include a visual breakdown of sums over $n$ order in Appendix~\ref{sec:bfemoments}. 
While the power is an efficient parametrization of the total system, streams are inherently a local (or restricted) measure of the potential, which we will explore below. 
Nevertheless, concisely describing the Milky Way--LMC system gives insight into the dynamics of the interaction.

The upper panel of Figure~\ref{fig:BFE_moment_evolution} demonstrates that the dipole is the largest deformation in the Milky Way. 
The magnitude of the dipole owes to the `stretching' of the Milky Way towards the LMC as the LMC approaches from beyond the virial radius. 
The dipole power becomes nonzero approximately as the LMC crosses the virial radius, and continues increasing until rising rapidly during the last few hundred Myr. 
Other harmonic orders are strongly subdominant, peaking below a tenth of the dipole power.

In contrast, the lower panel of Figure~\ref{fig:BFE_moment_evolution} shows that the quadrupole is the largest deformation of the LMC over the past 500~Myr, owing to the strong tidal forces experienced by the LMC near pericentre. 
While the strength of the quadrupole dominates the recent LMC response, other harmonic subspaces, including up to $l=6$, also contribute to the response. 
Prior to the past 500~Myr, the dipole deformation of the LMC also dominates, with nonzero power on a similar timescale as the Milky Way. 

In both panels, the estimation noise from finite-$N$ effects is visible in the coefficient series. 
This noise creates the smallest variations, most visible in the $l=1$ curves, with amplitudes of order $0.01\times10^4$ in the normalised power units.
However, there are larger amplitude variations as well (of order $0.05\times10^4$ in the normalised power units), that are likely the result of real dynamical evolution. 
As we cannot control the natural dynamical evolution of the models, we expect some secular evolution dynamics to also be encoded in the coefficient series.

\subsection{Density}\label{sec:BFE_dens}
\begin{figure*}
	\includegraphics[width=\textwidth]{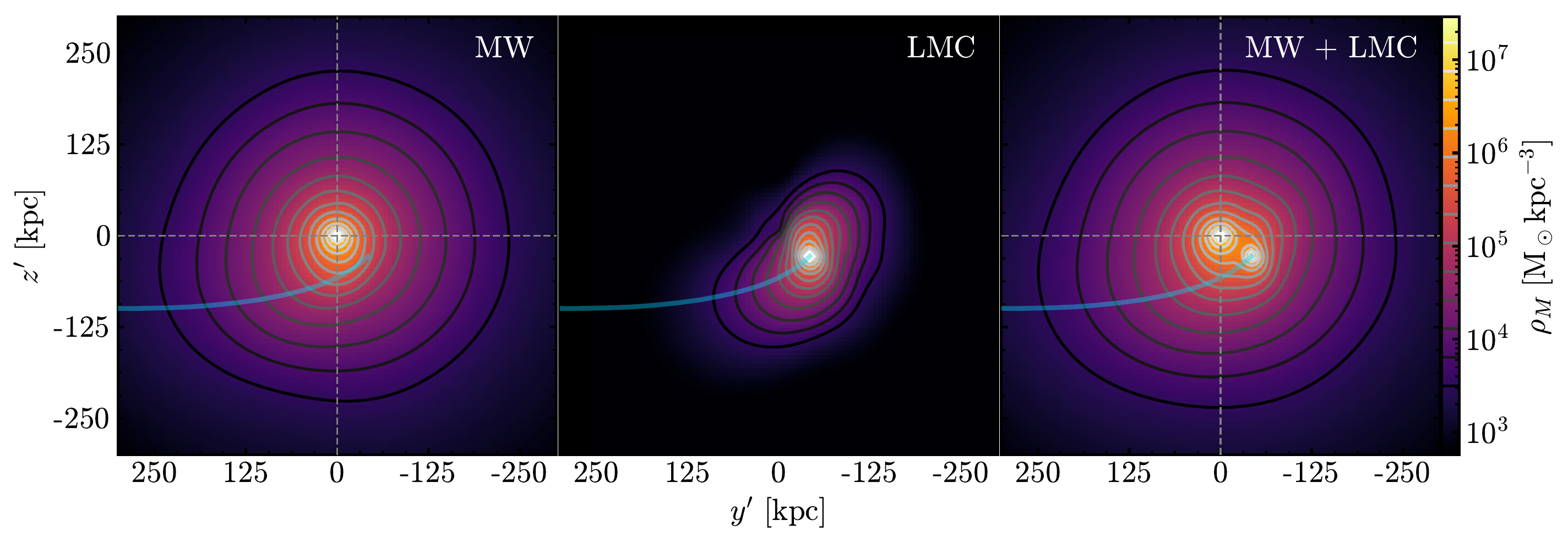}
    \caption{Density of the Milky Way halo (\textit{left panel}), the LMC (\textit{middle panel}), and combined (\textit{right panel}) of the BFE simulations in the orbital plane of the LMC at the present time. The density is projected over a 10~kpc thick slab. The contour lines show the densities with a constant multiplicative spacing of $\sim2$. The Milky Way exhibits deformations and a twist along the past orbit of the LMC (blue line). The dashed grey vertical and horizontal lines show the Milky Way centre to highlight asymmetries in the Milky Way halo. The LMC is heavily elongated along its orbit and twisted towards the centre of the Milky Way. A video of the LMC infall and the galaxies' deformations over the past 1.5~Gyr can be found \href{https://youtu.be/KMeGVMXfLTw}{here}. 
    }
    \label{fig:BFE_density}
\end{figure*}
We can use the basis function expansions to investigate how the Milky Way and the LMC deform in response to one another. 
We expect the biggest effect of deformations due to the LMC in its orbital plane since this is the plane in which the LMC's material spreads out the most \citep[e.g. see figure 10 in][]{Erkal..Orphan..2019}. 
In order to best show the LMC's effect, especially when it is far from the Milky Way, we rotate coordinates to be aligned with the LMC's orbital plane; 
this rotation is explained in Appendix~\ref{app:rot_mat}. 
While the LMC's orbital plane is very close to the Galactic $yz$-plane, it differs by $5.8^\circ$ in the $x$-direction, $4.3^\circ$ in the $y$-direction, and $3.9^\circ$ in the $z$-direction. 
\par Figure~\ref{fig:BFE_density}\footnote{Video link: \href{https://youtu.be/KMeGVMXfLTw}{https://youtu.be/KMeGVMXfLTw}} shows the densities of the Milky Way halo, the LMC halo, and their combined densities from the full BFE expansions. 
While their dark matter haloes are initialised to be spherical, over time (see video link in Figure~\ref{fig:BFE_density} caption), the Milky Way deforms where the LMC is falling in. 
This leads to a twist along the LMC's past orbit. 
The general shape of the Milky Way is governed by its dipole (as seen in Figure~\ref{fig:BFE_moment_evolution} and Figure~\ref{fig:app:ncoefs}).
The Milky Way's density is lopsided and the density contours are compressed in the positive $z'$-direction and expanded in the opposite direction, shifting the Milky Way's density downwards beyond $\sim 25$ kpc.

\par The LMC's first visible deformation is the quadrupole in the leading part of the LMC.
Around 500~Myr ago, with the increase of power in the quadrupole, the whole LMC starts elongating along the direction of its orbit.
Over the last 100~Myr, the inner LMC twists towards the Milky Way, which is described by the higher radial terms of the quadrupole (see the third row in Figure~\ref{fig:app:ncoefs}).
The LMC has some deformations, particularly towards the Milky Way, that need the higher harmonic orders to be described. 
These densities for both the Milky Way and the LMC are similar to other simulations of this interaction, e.g. in \citet{Erkal..Orphan..2019} and in \citet{GaravitoCamargo...BFE...21}.

\subsection{Forces}\label{sec:BFE_forces}
While the density is useful for highlighting the deformations, the density itself is not directly observable. Instead, the aim of this work is to show that the force fields generated by these deformations can be detected with stellar streams. In order to showcase how the deformations affect these forces, we consider several different expansions of our haloes throughout this paper. First, we consider the present-day monopole which captures the spherical behaviour of the halo. This is useful for comparing our deforming models to other modelling techniques used to represent the Milky Way and LMC which assume the Milky Way and LMC do not deform \citep[e.g.][]{Erkal..Orphan..2019, Vasiliev...tango3....2021, Shipp...LMCmassS5...2021}. We compare this with the `live' model which includes all orders of the multipole expansion.

Using the BFEs, we evaluate the forces from the dark matter haloes of the Milky Way and the LMC for the monopole and the live simulations. Since the monopole haloes are spherical, any aspherical forces in the live haloes must result from the deformations. The strength of the aspherical force is given by
\begin{equation}
    {F}_\mathrm{aspherical} = |\hat{r}\times\vec{F}|
\end{equation}
and the strength of the radial force is given by 
\begin{equation}
{F}_\mathrm{radial} = |\hat{r}\cdot\vec{F}|
\end{equation}
where $\hat{r}$ is the normalised position vector relative to the centre of the galaxy we considered.
We take the absolute values of these forces to understand the magnitude of the effect of the deformations.

\begin{figure*}
	\includegraphics[width=\textwidth]{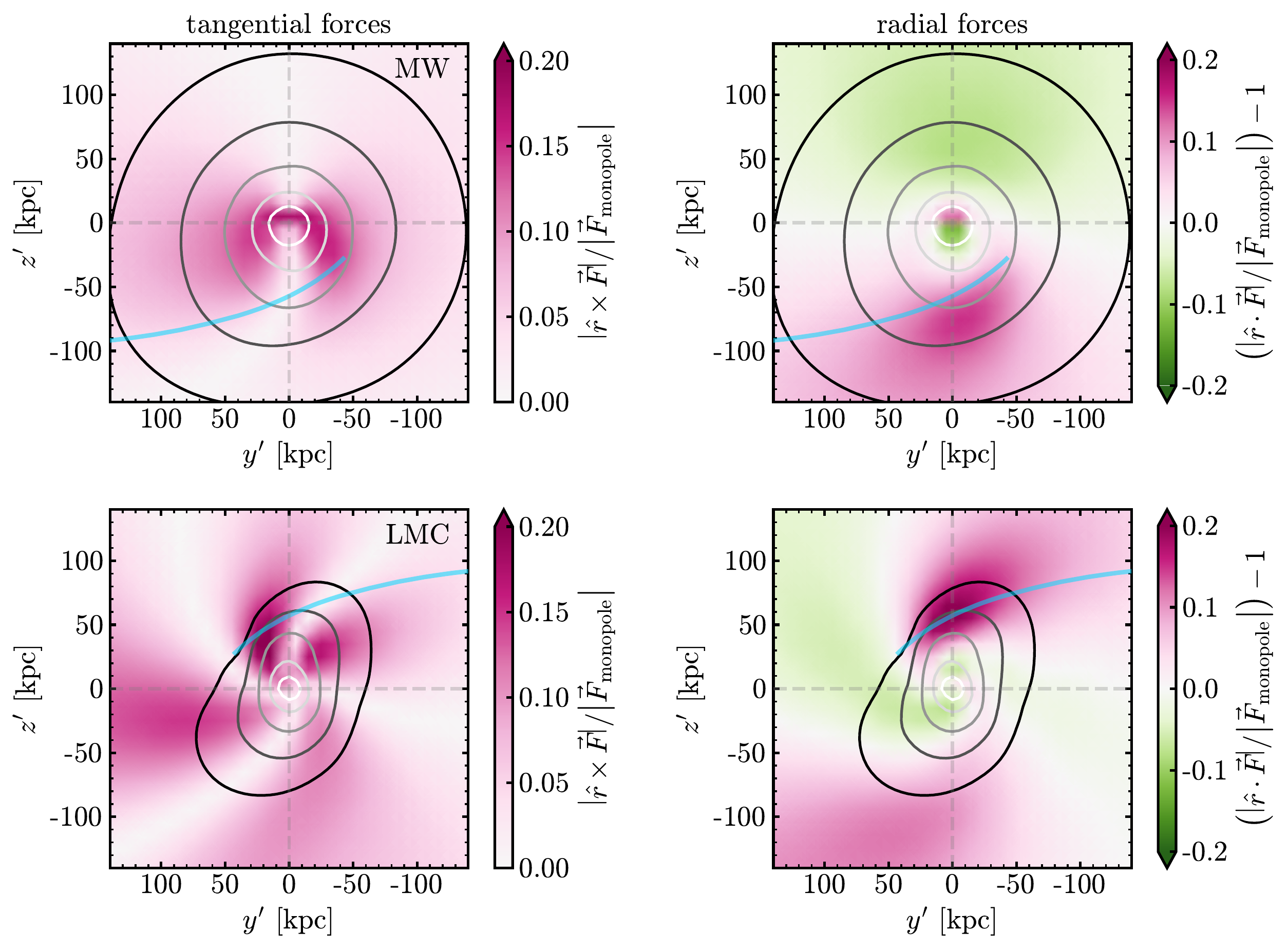}
    \caption{Aspherical and radial force differences between the live and monopole haloes for the Milky Way (\textit{top row}) and LMC (\textit{bottom row}) in the orbital plane of the LMC at the present time. To aid the comparison, we normalize the force differences by the force from the monopoles at the present day. The orbit of the other galaxy is indicated by the blue line. The grey dashed lines show the centres of each halo. The grey-scale contours show the halo densities with a constant multiplicative spacing of $\sim 3.8$. \textbf{Top left panel:} For the Milky Way, the aspherical forces are up to 18~per~cent of the monopole force and therefore could have a significant impact on objects in the affected parts. The maximum aspherical forces are around the centre of the Milky Way, oriented towards and away from the LMC, again showing the strong dipole. \textbf{Top right panel:} The radial forces in the North are lower than in the monopole forces. In the South, the radial forces are higher than the monopole forces, particularly following the LMC and its past orbit. Its change of force is higher compared to the Northern part. The first-order symmetry in both panels is due to the Milky Way's dipole. \textbf{Bottom left panel}: The aspherical forces of the LMC are split into four parts, indicating the prominent quadrupole. The borders of the quadrupole force shapes lie along the long and the short axis of the LMC's density distribution. The strongest parts of the aspherical force are at the leading part of the LMC, closest to the Milky Way, until $\sim 300$~Myr ago. Since then, the force field on the part opposite the Milky Way grew stronger. The aspherical forces of the LMC are up to 23~per~cent of its current monopole's force. \textbf{Bottom right panel:} The quadrupole of the LMC is visible in the radial forces as well. The strongest change in the radial forces is again where the Milky Way falls in. Very recently, the opposite direction gains radial force as well. The losses in the other two parts of the quadrupole are not as strong. A video of the infall centred on each galaxy and the galaxies' force fields over the past 0.7~Gyr can be found \href{https://youtu.be/im93oX6O33s}{here}.
    }
    \label{fig:BFE_force}
\end{figure*}

In Figure~\ref{fig:BFE_force}\footnote{Video link: \href{https://youtu.be/im93oX6O33s}{https://youtu.be/im93oX6O33s}}, we compare the aspherical and radial forces in the live and monopole haloes. In order to aid the comparison, we normalize the difference in these forces by the force from the monopole halo as a function of position. There are significant aspherical forces, up to 18~per~cent and 23~per~cent of the monopole forces of the Milky Way and LMC, respectively. The LMC experiences enhanced aspherical and radial forces, particularly at its leading arm for the last $\sim300$~Myr. While the Milky Way's dipole and the LMC's quadrupole are visible throughout the whole evolution, they become particularly significant over the last 100~Myr. Any objects that move in the areas of enhanced aspherical forces should be affected by the deformations. We also notice features in the Milky Way at small radii ($r<20~\kpc$) that likely correspond to the interaction of the disc and inner halo. Furthermore, over the course of the interaction and as seen in the top right panel of Figure~\ref{fig:BFE_force}, the Milky Way halo gets more centrally concentrated. For the purposes of this work, the small-scale features do not figure into the results. To understand the effects of the large-scale deformations better, in the next Section, we evolve and analyse a stellar stream in this time-dependent potential.

\section{The OC stream in live potentials}\label{sec:streams}
In this Section, we explore how the deforming Milky Way and LMC dark matter haloes affect the OC stellar stream \citep[e.g.][]{Grillmair...Orphan...2006, Belokurov...streams...2006,newberg+2010,Shipp...DES...18,koposov19}. We make this choice because the OC stream is one of the streams most strongly perturbed by the LMC \citep[e.g.][]{Erkal..Orphan..2019,Shipp...LMCmassS5...2021} and because its observed extent covers a wide range of radii in the Milky Way. As a result, we stress that we are only exploring the effect of the deformations on the small set of orbits that the OC stream inhabits as a demonstration of what is possible. Streams and structures on other orbits will also be strongly affected by these deformations and we will explore the effects on other streams in future work.

\subsection{Stream data}\label{sec:streams:data}

With proper motions from \textit{Gaia} EDR3 \citep{gaia_edr3}, radial velocities from the Southern Stellar Stream Spectroscopic Survey  \citep[\textit{S}$^5$;][]{li_S5}, and distances from RR Lyrae \citep[e.g.][]{koposov19}, we now have a 6D view of the OC stream. 
The presentation and detailed analysis of these data is published in \citet{Koposov2022}. The measurements of various observables are done through modelling of unbinned data as a function of angle along the stream track by cubic splines, similar to \citet{Erkal2017} and \citet{koposov19}. The number of likely spectroscopic members, from which the proper motions and radial velocities are measured, is 379. The number of likely RR Lyrae stars, from which the distance modulus is measured, is 120. 
The splines are specified in terms of their values at a sequence of knots. 
In Figure~\ref{fig:streams:full_live_stream}, we show the values at the knots with grey error bars. 
The stream coordinate system ($\phi_1, \phi_2$) is defined in \citet{koposov19} and the rotation matrix is given in their Appendix B.
Since there is little covariance between neighbouring data points, we treat these as independent measurements of the OC stream observables.    

\subsection{Stream modelling}\label{sec:streams:technique}
\par To model the OC stream, we use the modified Lagrange Cloud Stripping (mLCS) originally developed in \citet{Gibbons...mLCS...2014} which was further modified in \citet{Erkal..Orphan..2019} to include the force from the LMC and the reflex motion of the Milky Way. The progenitor is modelled as a Plummer sphere with an initial mass of $10^7\msun$, approximately matching the observational mass constraint from \citet{koposov19}, and a scale radius of 1~kpc, to roughly match the width of the OC stream \citep{Erkal..Orphan..2019}. We rewind the progenitor in the combined presence of the Milky Way and LMC for 4~Gyr. The orbital period of our OC stream model is 1.15~Gyr. The system is then evolved forwards and tracer particles are released from the progenitor's Lagrange points to generate a stream. The Lagrange radius is estimated by computing
\begin{equation}
r_t = \Bigg( \frac{G M_{\rm prog}}{\Omega^2-\frac{d^2 \Phi}{dr^2}} \Bigg)^\frac{1}{3}
\end{equation}
where $M_{\rm prog}$ is the mass of the progenitor, $\Omega$ is the angular velocity of the progenitor relative to the Milky Way, and $\frac{d^2 \Phi}{dr^2}$ is the second derivative of the Milky Way potential computed along the radial direction \citep{King+1962}. These ejected particles feel a force from the Milky Way, LMC, and progenitor. Due to the low mass of the progenitor, we do not model its dynamical friction in the presence of the Milky Way.

We model the Milky Way and LMC with the BFE potentials described in Section~\ref{sec:BFEs}. We evaluate the forces of each expansion (Milky Way halo, Milky Way stellar component, LMC halo) at each timestep for each particle. Motivated by the results of \cite{Dehen_Read_2011}, we compute two separate time scales for each particle. To capture the orbit around the Milky Way, we compute $\Delta t_{i,\,\rm orbit} = \eta \sqrt{\frac{r_i}{|\mathbf{a}_i|}}$ where $i$ is an index for the stream particles, $r_i$ is the distance to the Milky Way, $\mathbf{a}_i$ is the acceleration experienced by the particle due to the Milky Way, and $\eta=0.01$. To capture the orbit around the progenitor, we compute $\Delta t_{i,\,\rm prog} = \eta \sqrt{\frac{r_{i,\, \rm prog}}{|\mathbf{a}_{i,\,\rm prog}|}}$ where $r_{i,\,\rm prog}$ is the distance to the progenitor and $\mathbf{a}_{i,\,\rm prog}$ is the acceleration due to the progenitor. We then compute the minimum timestep over all of the particles, $\Delta t = \min\limits_{i} \left( \Delta t_{i,\,\rm orbit}, \Delta t_{i,\,\rm prog}\right)$ with a minimum timestep of $0.5$~Myr. The timesteps range up to 3~Myr. This is the same time-step criterion as used in \cite{Erkal..Orphan..2019}. We note that also including a time-step criterion for the orbit relative to the LMC makes no observable difference to the stream. 

Due to the long computation time of the force evaluations for each particle relative to an analytic force \citep[i.e. as in][]{Erkal..Orphan..2019}, we do not attempt to fit the data with these stream models. Instead, we find the best model from a grid search over the parameter space around the initial conditions for the progenitor in a spherical Milky Way halo with reflex motion from table~A1 in \citet{Erkal..Orphan..2019}. The streams created with these initial conditions are compared to the data in each observable (i.e. track, proper motions, distance, and radial velocity) following \citet{Erkal..Orphan..2019}. We use a right-handed coordinate system with the Sun's position at $x_\odot=(-8.249, 0, 0)\kpc$ \citep[distance to the Galactic center from][]{GravityCollaboration2020} and take its velocity to be $v_\odot = (11.1,245,7.3)\kms$ (with the peculiar velocity from \citealt{Schonrich2010} and basing the rotation velocity on proper motion measurements of Sgr\,A$^\star$ \citealt{Reid2004}). For each data point $i$, we select all model particles within $2^\circ$ in $\phi_1$. We fit a line as a function of $\phi_1$ to these particles. This maximum-likelihood fit returns the mean $m_{i\mathrm{, mod}}$ and uncertainty on the mean $\sigma_{i\mathrm{, mod}}$ at the location of the data point $i$. 
The uncertainty on the mean results from the finite number of particles in the stream model. We compare these fits to the data points in each observable using the log likelihood\footnote{Where $\log$ denotes the natural logarithm.} 
\begin{equation}
    \log\mathcal{L}_i = - \frac{1}{2} \log \left( 2\pi \left( \sigma_{i\mathrm{, obs}}^2 + \sigma_{i\mathrm{, mod}}^2 \right) \right) - \frac{1}{2} \frac{\left( m_{i\mathrm{, obs}} - m_{i\mathrm{, mod}}\right)^2}{ \sigma_{i\mathrm{, obs}}^2 + \sigma_{i\mathrm{, mod}}^2}
\end{equation}
with the observed values $m_{i\mathrm{, obs}}$ and observed uncertainty $\sigma_{i\mathrm{, obs}}$. We sum these log-likelihoods together and select the stream with the highest likelihood as the best-matching stream. We use these initial conditions for all stream models shown in this paper. These initial conditions for the stream progenitor are $\phi_1=6.340^\circ$, $\phi_2=-0.456^\circ, d=18.975\kpc$, $v_r=93.786\kms$, $m_\alpha^\ast=-3.590~{\rm mas~yr}^{-1}$, $m_\delta=2.666~{\rm mas~yr}^{-1}$, following the notation of \citet{koposov19} and \citet{Erkal..Orphan..2019}.
The stream track coordinates ($\phi_1, \phi_2$) are given in a coordinate system aligned with the OC stream provided in \citet{koposov19}.
This coordinate system follows a great circle with a pole at $(\alpha_{\rm OC}, \delta_{\rm OC}) = (72^\circ, -14^\circ)$, and has its origin at $(\alpha_0, \delta_0) =(191.10487^\circ, -62.86084^\circ)$. 
The rotation matrix to this coordinate system is given in Appendix B of \citet{koposov19}.

\begin{figure}
	\includegraphics[width=\columnwidth]{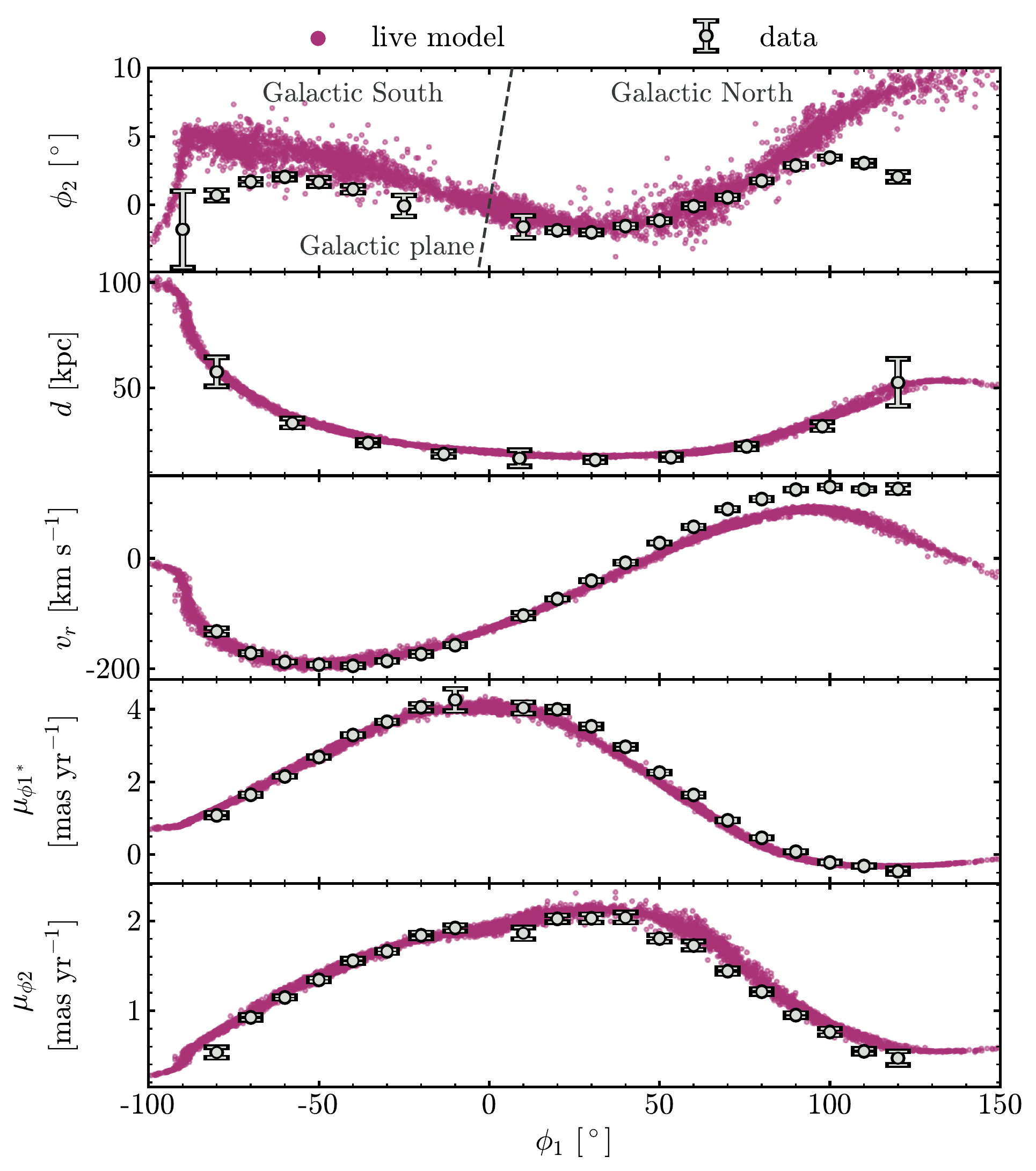}
    \caption{Observables of the best-matching OC stream model and data. The rows show the stream in sky coordinates, heliocentric distance, radial velocity in the Galactic standard of rest, and proper motions in stream coordinates,  not reflex-corrected, respectively. The grey points with error bars show the observed stream from \citet{Koposov2022}, and the purple points are the simulated stream particles of the best-matching stream in the fully evolving Milky Way--LMC simulations. The model stream matches the trends of the observed stream well but there are some quantitative differences. }
    \label{fig:streams:full_live_stream}
\end{figure}

The stream observables of the best-matching stream are shown in Figure~\ref{fig:streams:full_live_stream}. The model follows the overall trends of the observed stream well but there are quantitative discrepancies, in particular the radial velocities in the Northern part of the stream ($\phi_1 > 0^\circ$). We note that \citet{Erkal..Orphan..2019} did not use radial velocities when fitting the stream. Since our simulated Milky Way and LMC are initialized with the potentials \citet{Erkal..Orphan..2019} obtained from their fits, a grid around their best initial conditions therefore might limit our capacity to fully match the radial velocities. Furthermore, including radial velocities could yield different potentials in which we would match the data better.
This may explain part of the discrepancy. 
In addition, our potentials evolve due to the interaction of the Milky Way and LMC. As a consistency check, we let the stream evolve in the initial potential with the initial conditions of \citet{Erkal..Orphan..2019} and find that the stream matches the stream track, distance, and proper motions as in \citet{Erkal..Orphan..2019} which is much better than the live model presented here. However, for the scope of this work, the live model stream resembles the OC data close enough.
We stress that this paper does not aim at fitting the data, but rather at investigating how the best-matching OC stream model is affected by the deformations of the Milky Way and the LMC. 

\subsection{Stream tracks in increasingly complex potentials}\label{sec:streams:tracks}
\begin{figure*}
    \centering
    \includegraphics[width=\textwidth]{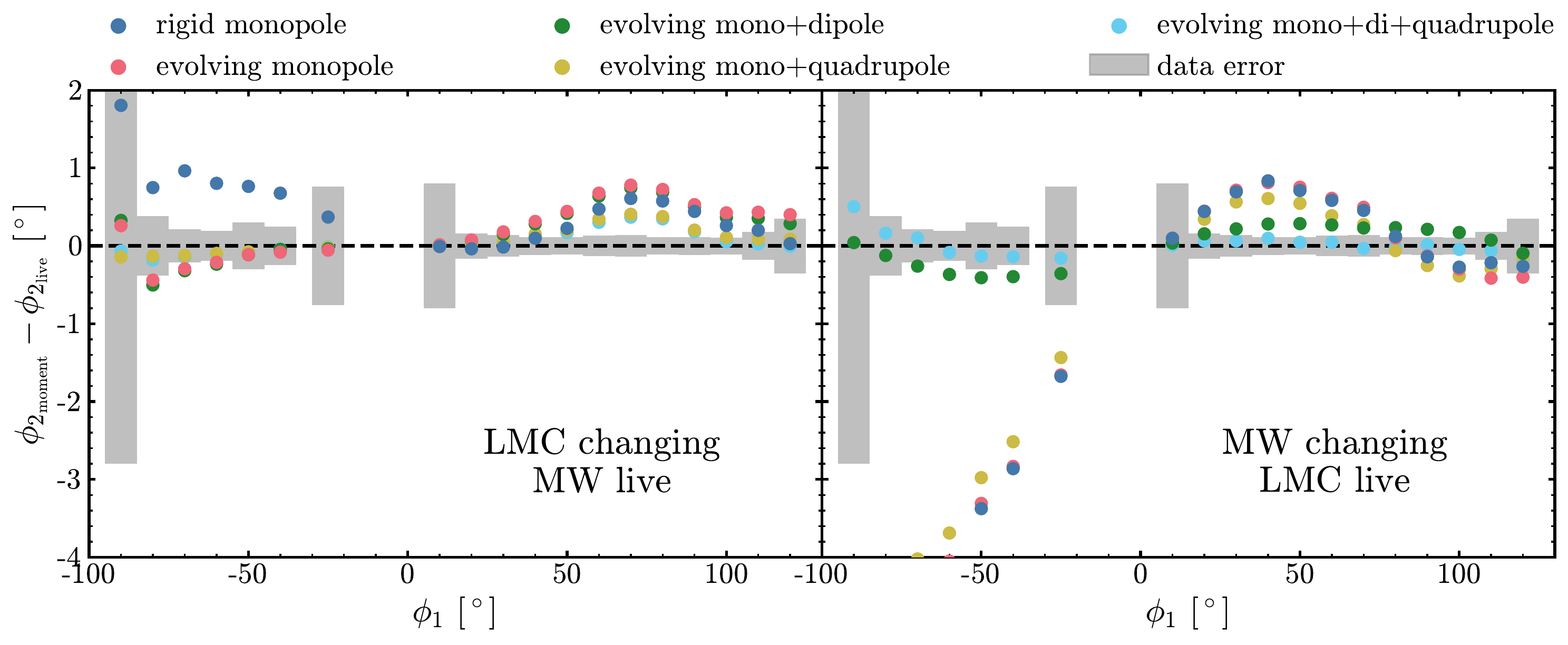}
    \caption{Tracks of the simulated OC stream in different LMC and Milky Way potentials compared to the fully live potential. 
    In the left plot, the Milky Way is live and different moments of the LMC are selected as the potential.
    In the right plot, the LMC is live and we vary the Milky Way. 
    All potentials include the live Milky Way disc.
    The moments are the rigid monopole (blue), evolving monopole (red), then, in addition to the monopole, the dipole (green) and quadrupole (yellow). 
    Finally, we combine monopole, dipole and quadrupole (cyan).
    The grey bars show the size of the uncertainty on the mean ($\pm 1 sigma$) for the observed stream track.
    All stream tracks are binned as explained in Section \ref{sec:streams:technique}, at the same positions in $\phi_1$ as the data.
    The errors of the mean observable tracks of the model, which depend on the number of particles in the model, are smaller than the size of the markers, and therefore not shown.
    For the LMC, the strongest effect results from turning on the time dependence (blue to red).
    From the higher moments, the quadrupole determines most of the live track (red to yellow). 
    In the Northern part, between $\sim 50^\circ < \phi_1 < 80^\circ$, the live track indicated by the black dashed line differs from the others by a significant amount.
    The Milky Way's most important moment is the dipole (red to green).
    The tracks of the most affected configurations are shown in the top panels of Figure~\ref{fig:app:streams:observables}.
    Some of these effects, in particular the Milky Way dipole's effect, are much greater than the observed errors.}
    \label{fig:streams:track_oc}
\end{figure*}
\par One of the advantages of BFEs is that for each expansion, we can select different moments to evaluate the contribution of different harmonic orders and radial subspaces. 
In order to better understand how the different terms of the BFEs affect the stream, we now selectively turn off the contribution from certain functions in the total expansion. 
To see how the inclusion or exclusion of these moments affects the OC stream, we run the model with the same initial conditions for the progenitor in the different potential setups. 
Figure~\ref{fig:streams:track_oc} shows these different OC stream models.
In order to isolate the effect of each term in the expansion, we keep one galaxy live, while we change the moments of the second galaxy, from a rigid monopole to the evolving lower order moments, to the fully live case. We note that the Milky Way disc is kept live in all cases.

\par First we consider the monopole which describes the spherical shape of each galaxy. 
Since the monopole is time-dependent, we first consider the final monopole at the present day (i.e. $t=0$~Gyr) which we dub the `rigid monopole'. This is to be contrasted with the `evolving monopole' which includes the monopole's time dependence. The OC stream in the rigid monopole and evolving monopole are shown in blue and red respectively in Figure~\ref{fig:streams:track_oc}. Interestingly, using a rigid monopole for either the Milky Way or LMC results in a significantly different stream than the full live stream. Furthermore, including the evolving monopole does not fully remedy this, showing that the higher order terms are also important. The other moments shown are the evolving dipole, quadrupole, and dipole+quadrupole together in green, yellow, and cyan, respectively. For all higher moments, the evolving monopole is included so that the additive effect of each component can be seen. The `evolving mono+di+quadrupole' model captures all effects with $l \leq 2$. Given current observational uncertainties, it is indistinguishable from the live model, suggesting that higher order terms (i.e. $l\geq3$) would not be necessary to describe the deforming Milky Way.

For the LMC, the biggest effect on the Southern part of the OC stream is from turning on the time-dependence of the monopole (blue to red), followed by the quadrupole (red to yellow). In the Northern part of the stream, between $\sim 50^\circ < \phi_1 < 80^\circ$, there is an offset between the `evolving mono+di+quadrupole' and the live model, showing that higher order moments are playing a significant role.

For the Milky Way, the biggest impact comes from including the dipole (i.e. green and cyan points), showing the impact of the aspherical forces. The forces from the dipole pull the OC stream up in the South and down in the North. This effect is larger than the effect from any orders of the LMC's expansion, showing that accounting for the Milky Way's dipole deformation is the next most important effect to consider for the OC stream. We stress that the differences between the moments are much larger than the uncertainty in the data which shows that fits to the OC stream should be sensitive to these deformations. We note that we have focused on changes in the stream track since these produce the largest effect compared to observational uncertainties. We show the differences in all observables in Appendix~\ref{app:all_observables} for comparison.

\par While this shows how different moments affect the stream track, it has an important caveat. When we change the BFE, the orbit of the progenitor also changes. This might affect the resulting stream tracks which complicates the comparison of the different expansion orders. To make a cleaner comparison, we next investigate the contribution of each moment of each galaxy towards the forces experienced by the live stream.

\subsection{Integrated absolute forces}\label{sec:streams_forces}
In the previous Section, we explored how the different components of the BFE affect the stream track with the understanding that their orbits slightly differ. To more directly explore the impact of each component, we now compare the integrated absolute forces from the different moments on the same stream.

In order to dissect the contributions of each moment to the total force each stream particle experiences, we first run our best model stream in the fully live potential. We save the forces due to each moment from each galaxy on each stream particle every 20~Myr. We then sum the absolute values of these forces over all the snapshots to get the integrated absolute force on each particle: 
\begin{equation}
    F_{\sum\mathrm{tot}} = \sum_t|\vec{F}|\mathrm{d}t .
\end{equation}
We also include the integrated absolute force on the progenitor up to the time when each particle was stripped to account for the forces the particles experienced before they left the progenitor. We consider different components of the force to better understand how the stream is affected. The total force is the force each particle experiences from each galaxy. The aspherical force is the force the particles experience from the non-spherical components,
\begin{equation}
    F_{\sum\mathrm{aspherical}} = \sum_t|\hat{r}\times\vec{F}|\mathrm{d}t.
\end{equation}
As discussed in Section~\ref{sec:BFE_forces}, the Milky Way and the LMC haloes are initially spherical and do not exhibit aspherical forces. Any aspherical forces come from deformations and therefore these are an indication of how dark matter deformations affect the OC stream.
\begin{figure*}
    \centering
    \includegraphics[width=\textwidth]{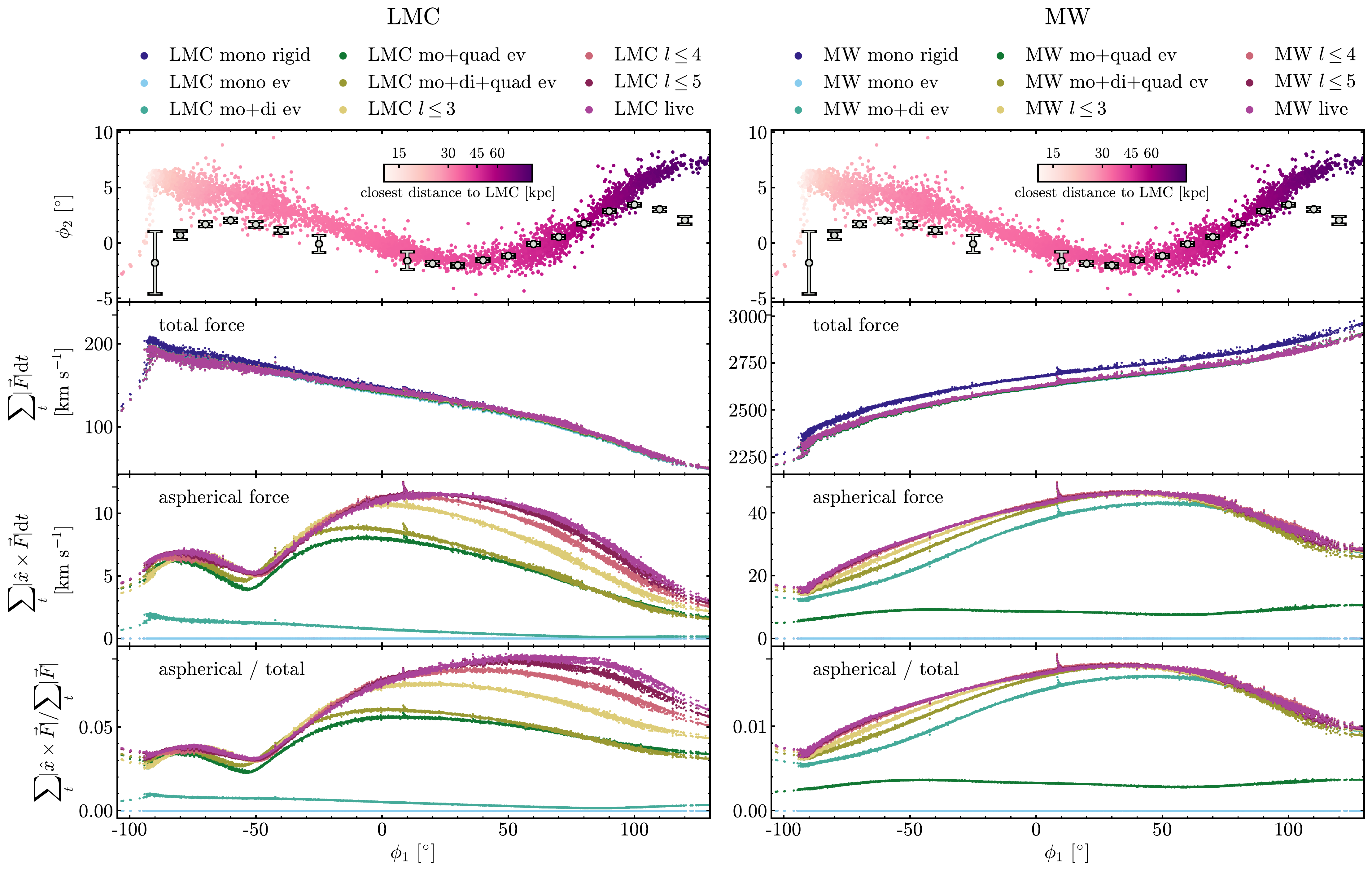}
    \caption{Integrated absolute forces of different moments of the LMC (\textit{left}) and the Milky Way (\textit{right}) on each particle of the stream. The \textbf{top row} shows the stream track, colour-coded by the closest approach distance to the LMC. The Southern portion of the stream ($\phi_1<0^\circ$) passes the closest to the LMC, within 15-30~kpc. The \textbf{second row} shows the total integrated absolute forces. The colours indicate the moments whose forces are evaluated, going from blueish colours for the spherical components to reddish colours for the live potentials. Between the different moments, there is almost no difference. The total force of the Milky Way on the stream is more than a magnitude higher than the LMC's force since the Milky Way is more massive and since the OC stream orbits the Milky Way during the whole integration time ($4~\mathrm{Gyr}$) while it is only impacted by the LMC over the most recent several hundred Myr. The \textbf{third row} shows the integrated aspherical forces. These forces are particularly interesting since these can only be a result of deformations. The moments with the highest marginal change are the quadrupole (green) for the LMC and the dipole (teal) for the Milky Way. In the LMC, the quadrupole contributes to most of the live forces on the Southernmost part of the stream ($\sim\phi_1<-40^\circ$). The Northern part of the stream is affected by higher moments beyond the quadrupole to build up to the live forces. In the Milky Way, the dipole and then quadrupole make up the majority of the live forces. The apparent spike at $\phi_1\sim10^\circ$ is caused by particles that have just left the progenitor. The \textbf{bottom row} shows the ratio of aspherical to total integrated absolute forces. For the Milky Way, the aspherical forces are up to 1.8~per~cent of the total forces while for the LMC they are up to 10~per~cent.}
    \label{fig:streams:f_int}
\end{figure*}

The forces from the LMC and the Milky Way are shown in Figure~\ref{fig:streams:f_int}. The total force of the LMC is higher on the stream particles that passed closer to the LMC and roughly similar for all potential setups. The Milky Way exerts more force on the Northern part of the stream because it is the leading arm of the stream and hence has a smaller pericenter and thus a larger force. Since the monopoles only exert radial forces, the aspherical forces can only be due to deformations of the haloes and are thus more instructive for comparing the different expansions. The moments contributing most to the live force are the dipole for the Milky Way and the quadrupole for the LMC. Interestingly, for $\phi_1 > -45^\circ$, the aspherical forces exerted by the LMC need the contribution of all higher moments to build up to the live forces showing that high multipole orders ($l \geq 3$) are needed to adequately capture the LMC's effect.

\par The ratio of the aspherical forces to live forces is up to 10~per~cent for the LMC while only up to 1.8~per~cent for the Milky Way. We stress that this does not directly translate to the impact of the deformations on the stream since the OC stream orbits the Milky Way throughout the whole integration time (4~Gyr). However, the Milky Way has only been deforming for the last $\sim 500$ Myr (see e.g. Figure~\ref{fig:BFE_moment_evolution}). Therefore, the OC stream has experienced only radial forces for 3.5~Gyr and they make up most of the integrated absolute forces. If we restrict to the most recent 1 Gyr,  the ratio of aspherical to total forces for the Milky Way reaches 5.5~per~cent. In contrast, the LMC is already deforming when it gets close to OC so this metric should capture the relative strength of the aspherical forces. We discuss this in more detail in the next Section.
 
\section{Discussion}\label{sec:discussion}
\subsection{The effect of deformations on streams}\label{sec:disc:deformation_effect}

Due to its length, OC experiences different effects from the deformations along the stream at different times. 
\begin{figure}
	\includegraphics[width=\columnwidth]{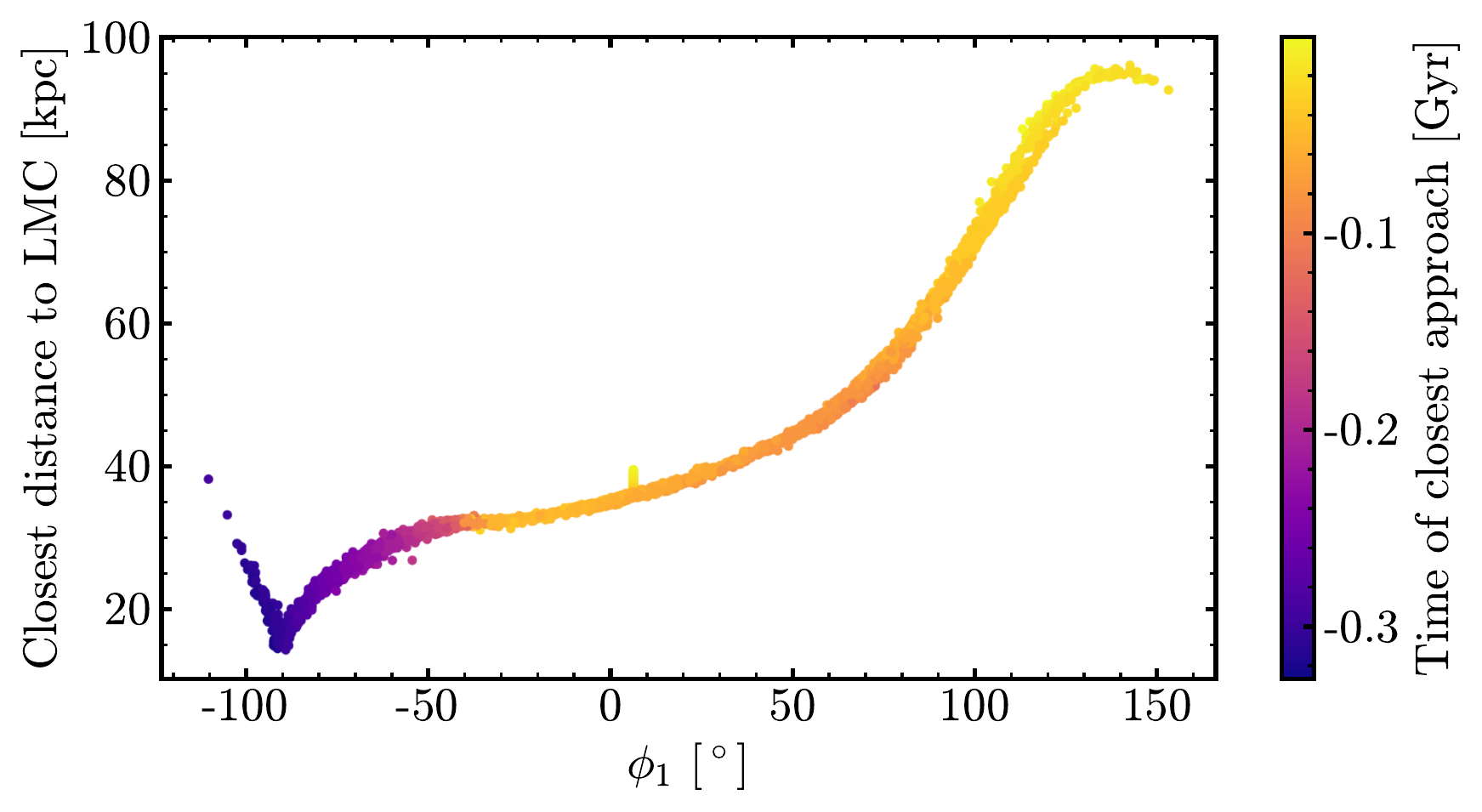}
    \caption{Distance and look-back time of the closest approach of the OC stream particles and the LMC. The plot shows the closest distance to the LMC along the stream colour-coded by the look-back time when this approach happened. The Southernmost part $\left(\phi_1<-45^\circ\right)$ came closest to the LMC ($10-30\kpc$), around $200-300$~Myr ago while the rest of the stream had its closest interaction more recently ($<100$~Myr ago) and experienced a more strongly deformed LMC. }
    \label{fig:disc:OC_LMC_d_t}
\end{figure}
In Figure~\ref{fig:disc:OC_LMC_d_t}, we show the distance and time of the closest approach for each stream particle. While the Southernmost part of the stream passed by the LMC around 300~Myr ago, the rest of the stream had its closest passage only within the last 100~Myr. There is a sharp cut between these time scales at $\phi_1=-45^\circ$. At this point in the stream, there is a dip in the aspherical forces experienced by the stream due to the LMC (see Figure~\ref{fig:streams:f_int}). The aspherical forces on the left side of this dip are governed by the quadrupole which started growing around the time that this portion of OC had its closest passage to the LMC ($\sim$300~Myr ago, see Figure~\ref{fig:BFE_moment_evolution}).

The particles with $\phi_1>-45^\circ$ had their closest approach to the LMC recently, and at a greater distance than the Southernmost part of the stream. Over the last 100~Myr, the quadrupole of the LMC increased significantly, but so did the higher moments (see Figure~\ref{fig:BFE_moment_evolution}). These particles experienced an increased effect, not only by the quadrupole but also by the higher terms that allow for deformations on smaller scales (see bottom two rows of Figure~\ref{fig:app:ncoefs}). We observe this effect in the aspherical forces from the LMC on the stream in Figure~\ref{fig:streams:f_int}. While the quadrupole still makes up approximately half of the summed absolute aspherical forces, the contributions of all higher moments are necessary to get to the absolute sum of aspherical forces on the live stream.

\cite{Shipp...LMCmassS5...2021} have modelled the OC stream in the presence of the LMC with rigid Milky Way and LMC models. Interestingly, at the position of the dip in the aspherical forces $\left(\phi_1=-45^\circ\right)$, \cite{Shipp...LMCmassS5...2021} find a sudden change in the orientation of the LMC's perturbation on the OC stream (see their figure~5). They approximate perturbations with velocity kicks split up into different components: along the angular momentum of the stream, and in the radial and tangential direction of the orbit of the stream. For the OC stream, the velocity kick in the angular momentum direction is the strongest along most of the stream, but at the position of the dip, the offset in angular momentum has a sharp dip while the tangential offset has a quick rise. Moreover, their figure~8 shows the position of the stream with respect to the LMC at the time of the closest approach. The OC stream is split up into two unconnected parts at different positions and orientations. This is the same split that we see in the colours of Figure~\ref{fig:disc:OC_LMC_d_t}. These results from \citet{Shipp...LMCmassS5...2021} and our results show that different regions of the stream are sensitive to different parts of the LMC.

\subsection{Possible bias in Milky Way and LMC halo measurements}

In this work, we have shown that the deforming Milky Way and LMC haloes have a significant effect on the OC stream. These effects have been ignored in all previous stream fits which have been done with rigid haloes for the Milky Way and LMC \citep[i.e.][]{Erkal..Orphan..2019,Vasiliev...tango3....2021}. Interestingly, the change in the on-sky position from including the deformations ($\sim1^\circ$, see Figure~\ref{fig:streams:track_oc}) is similar to the change when allowing for the Milky Way halo to be flattened \citep[$\sim 0.5^\circ$, see figure~9 in][]{Erkal..Orphan..2019}. This suggests that at least some of this flattening could be due to deformations instead of the intrinsic shape of the Milky Way. This could be an explanation of the peculiar halo shapes inferred with the OC stream and Sagittarius stream which are strongly flattened and not aligned with the disc \citep[][]{Erkal..Orphan..2019,Vasiliev...tango3....2021}. Alternatively, if these measurements of the halo are robust, they could indicate that the Milky Way halo is twisted and aligned with the LMC's orbital plane as in \cite{Shao+21}. We will investigate these potential biases in future work by fitting the stream models generated in this work with the same rigid haloes that are used to fit real streams \citep[e.g.][]{Erkal..Orphan..2019}.

\subsection{Future directions}

Ultimately, we want to measure the Milky Way and LMC's time-dependent haloes by fitting streams in the Milky Way with models that include these effects. On the data side, we already have precise measurements for Sagittarius \citep[e.g.][]{ibata+2020,ramos+2020,Vasiliev...tango3....2021} and the OC stream \citep[e.g.][]{koposov19, Koposov2022}. These are the longest streams in the Milky Way that, along the stream, experienced different effects of the deformations at different times. Both have been used for fits of the Milky Way \citep[e.g.][]{Law...Majewski...2010}, recently including models of the LMC \citep[e.g.][]{Erkal..Orphan..2019, Vasiliev...tango3....2021}. There are also several shorter streams that have been affected by the LMC \citep{Shipp...DESpms...19}. \citet{Shipp...LMCmassS5...2021} have shown that the LMC has a significant impact on the OC stream and four additional streams in the Southern Galactic hemisphere, and has used them to fit the mass of the LMC. These streams had their last closest approach to the LMC between $\sim100$~Myr to $\sim10$~Myr ago, with closest approach distances ranging between $4-40$~kpc. As an ensemble, these streams pass through different parts of the deforming Milky Way and LMC at differing times and thus we should be able to use them to measure the time-dependence of the Milky Way and LMC haloes with upgraded BFE technology. 

To be able to fit these streams, we need interpolatable BFEs. For this, we need a set of simulations with different Milky Way and LMC masses and, ultimately, different initial halo shapes. With these simulations, we then need to be able to interpolate between the coefficients to vary all parameters and fit the haloes with stellar streams. At each timestep, these simulations have 2058 coefficients to describe the haloes and 234 coefficients to describe the disc. To be able to fit these potentials, we need to improve our understanding of the importance of different terms in the expansion. One promising avenue for understanding and decomposing the time-dependence of the basis function expansions is with multichannel singular spectral analysis \citep{Weinberg...MSSA...2021}. This non-parametric technique had some success isolating important dynamical effects but is still being developed.

\subsection{Implications for dark matter and alternative gravity models}
In this work, we have argued that the OC stream is sensitive to the deformed dark matter haloes of the Milky Way and LMC. Detecting these deformations would allow us to test a robust prediction of the collisionless nature of cold dark matter and would allow us to constrain alternative dark matter and alternative gravity models. 

In self-interacting dark matter (SIDM), the LMC and Milky Way dark matter haloes would experience additional forces during their interaction. Early work by \cite{Furlanetto+2002}, who approximated the SIDM as a perfect fluid, found that the structure of the wakes created in the dark matter haloes depends on whether the interaction is subsonic or supersonic. It remains to be seen what these effects look like in SIDM models with smaller cross-sections that do not assume a perfect fluid. We note that the Milky Way--LMC system will also be useful for constraining velocity-dependent SIDM models \citep[e.g.][]{ackerman+2009,kaplinghat+2016} since the velocities are much smaller than those of cluster-scale systems which are currently providing some of the most stringent limits \citep[e.g.][]{tulin+2018}. Similarly, the effect of dynamical friction and the gravitational wake created is different in fuzzy dark matter models \citep[e.g.][]{hui+2017,lancaster+2020}.

In alternative gravity models, e.g. modified Newtonian Dynamics \citep[MOND, e.g.][]{Milgrom_1983,Bekenstein+1984}, the outer Milky Way has no dark matter halo to deform but still has a stellar halo and a hot gaseous corona \citep[][]{gatto+2013,miller+2015}. Since the stellar halo and corona densities have a different radial dependence than the dark matter halo \citep[e.g.][]{miller+2015,xue+2015} they would presumably also deform differently. In addition, since dynamical friction behaves differently in the CDM paradigm and MOND \citep[e.g.][]{ciotti+2004,nipoti+2008}, the orbit of the LMC is quite different \citep[e.g.][]{Wu+2008,Schee+2013} which would also affect how the outer Milky Way deforms. As a result, we expect that detecting a deformed Milky Way potential with stellar streams would likely provide a powerful discriminator between dark matter and alternative gravity models. In order to confirm this, more work is needed to understand how the Milky Way deforms in alternative gravity models and how these deformations affect stellar streams.

\subsection{The effect of the Small Magellanic Cloud}

Our present study neglects the presence of the Small Magellanic Cloud (SMC) in our model of the Milky Way--LMC interaction. One can make estimates for the possible bias our models incur from this omission from simple BFE-based arguments. In essence, we ask: can the SMC be responsible for levels of deformation in the LMC that would be detectable in our expansion, particularly at spatial locations that would be important for the OC stream? While the SMC is likely responsible for the deformations observed in the luminous component of the LMC, the deformations in the LMC dark matter halo are likely dominated by the interaction with the Milky Way.

\par Following the LMC--SMC models developed in \citet{Cullinane_MagesII_2021}, we choose a mass ratio of $M_{\rm SMC}/M_{\rm LMC} = 2.5\times10^9\msun/7.17\times10^{10}\msun = 0.03$ to represent their present-day masses. If we assume that the LMC expansion absorbs the contribution from the SMC, the mass ratio projects immediately into the monopole term, increasing the monopole by the mass ratio fraction (i.e. making the LMC heavier). More interesting are the implications for the higher-order terms in the expansion. If we assume that the SMC is represented by a Hernquist spherical mass distribution with $r_s=10$~kpc centred on the current location of the SMC, the contribution to the amplitude of the dipole terms in the spherical expansion is of order 0.1~per~cent relative to the LMC, i.e. an order of magnitude smaller than the deformations induced by the Milky Way. In this simple scenario, the SMC does not contribute to even-order spherical harmonics, which are the most influential for the evolution of the OC stream. 

It is likely that the SMC had a higher mass in the past, but repeated interactions with the LMC have stripped much of the mass from the SMC \citep[e.g.][]{besla+2012,de-leo+2020}. This mass will be distributed around the LMC, with the primary contribution being to the monopole and likely having little impact on higher-order harmonics. Future models should investigate the possible geometries of stripped SMC dark matter mass around the LMC.

We note that this global view of the influence of the SMC on the LMC does not describe the forces orbits near the SMC might feel; thus we caution that for orbits that pass near the SMC at any point in the past (e.g. smaller than the LMC--SMC separation), the influence may be much larger \citep[e.g. SMC's effect on orbits of dwarfs;][]{Patel+2020,simon+2020}. In order to test this, we injected the SMC as a tracer particle in our simulation. We find that, compared to the LMC, it has a more distant approach to the OC stream, suggesting it will have a negligible effect. Along similar lines, \citet{Koposov2022} explore this in more detail by modelling the gravitational effect of the SMC on the OC stream and find that it has almost no effect on the stream.

\subsection{The complex landscape of the Milky Way potential}

Although we have focused on the interaction of the Milky Way and the LMC, there are many other effects which can perturb streams and other tracers in the Milky Way, and thus complicate our ability to measure the potential. For example, the Sagittarius dwarf was likely much more massive when initially accreted onto the Milky Way, $\gtrsim 6\times10^{10}\msun$ \citep[e.g.][]{Bland-Hawthorn+2021,Gibbons2017}. The stripped dark matter of Sagittarius could have a substantial effect on the potential as well as on streams that pass through it \citep{Bovy2016}. Similarly, the accretion of Gaia-Sausage-Enceladus \citep[GSE,][]{Belokurov+2018,Helmi+2018} likely deposited a substantial amount of dark matter in the inner halo. \cite{Han+2022} argue that this may cause a long-lived tilt in the dark matter halo that supports the orbits of stars in the Hercules Aquila Cloud \citep[e.g.][]{Belokurov+2007,Simion+2014} and Virgo Overdensity \citep[e.g.][]{Vivas+2001,Newberg+2002}. On slightly larger scales, \cite{Valluri+2021} have shown that the expected figure rotation of the Milky Way's dark matter halo can also affect streams. On smaller scales, many works have shown that the bar and spiral arms of the Milky Way can perturb streams \citep[e.g.][]{Hattori+2016,Price-Whelan+2016,Erkal2017,Pearson+2017,Banik+Bovy2019}.

Going forward, it will be crucial to identify which of this myriad of effects is the most important in each region of the Milky Way. Ideally, there will be a leading order effect in each region, e.g. the bar and spiral arms in the inner Milky Way, and the LMC in the outer Milky Way. This would allow us to search for and convincingly measure these effects without having to build overly complex models that contain all of these perturbations.

\section{Conclusions}\label{sec:conclusions}
The Milky Way and the LMC are in disequilibrium \citep[e.g.][]{GaravitoCamargo...DMwake...19, GaravitoCamargo...BFE...21, Erkal...MWbias...20, Erkal...MWbias2...21, Conroy...21, Petersen...reflexmotion...21, Petersen...strippedLMC...21}, and several stellar streams in the Milky Way halo are affected, particularly the OC stream \citep{Erkal..Orphan..2019, koposov19,Shipp...LMCmassS5...2021}. In this work, we have explored the effect of deformations of the dark matter haloes of these galaxies on a simulated OC stream using basis function expansions. We presented a $N$-body simulation of the interaction between the Milky Way and the LMC run with the BFE code \textsc{exp} \citep{weinberg99, petersen..exp..21}. This allowed us to evaluate the time-dependent forces at any position in the system, and thus evolve the OC stream in a deforming Milky Way and LMC system.
We compared our simulations to the exquisite 6-D phase-space map of the OC stream from \textit{Gaia} and the \textit{S$^5$} survey \citep{Koposov2022}.

Equipped with these BFEs, we investigated several aspects of the effect of the deformations on the OC stream. 
Our results are the following: 
\begin{itemize}
    \item The Milky Way deformations owe primarily to the dipole, which has grown in strength over the last $\sim500$~Myr. In contrast, the LMC deformations owe primarily to the quadrupole, which steeply gained power over the last $\sim200$~Myr. During the last $\sim100$~Myr, the LMC's higher moments have also gained some power (see Figure~\ref{fig:BFE_moment_evolution}). These multipoles are also visible in the density and force field plots and videos of the Milky Way and the LMC (see Figures~\ref{fig:BFE_density} and \ref{fig:BFE_force}).
    \item The OC stream changes significantly when evolved in the presence of different moments (and therefore different potentials) with the same initial conditions for the stream progenitor. These effects are much larger than the current observational uncertainties so models that ignore these effects may be biased. The biggest change is induced by the Milky Way's dipole, followed by the time-evolving monopole of the LMC and the LMC's quadrupole (see Figure~\ref{fig:streams:track_oc}).  
    \item In order to isolate the impact of each multipole, we evolve the stream in the live potential and calculate the force contributions, with a particular emphasis on the aspherical forces that are due to deformations (see Figure~\ref{fig:streams:f_int}). Again, the dipole contributes the most to the Milky Way force while the LMC forces are dominated by the quadrupole. Interestingly, for the Northern part of the OC stream, higher moments of the LMC expansion ($l\geq 3$) make a significant contribution to the force it exerts on the OC stream.
    \item The OC stream is sensitive to the time-dependent deformation of the LMC. This is due to the fact that different parts of the OC stream experience their closest passage with the LMC at different times (Figure~\ref{fig:disc:OC_LMC_d_t}), and that the high-order multipoles of the LMC are growing over the past $\sim 100$~Myr (Figure~\ref{fig:BFE_moment_evolution}). As a result, the Northern and Southern components of the OC stream experience different force contributions from the multipole orders (see Figure~\ref{fig:streams:f_int}).
\end{itemize}

In summary, we have shown that the OC stream acts as a sensitive detector in the Milky Way--LMC dark matter collider. If our models are correct, these effects are already present in the data since they are much larger than the observational uncertainties of the OC stream. Progress is now needed on the theory side to fully utilize this data to measure the time-dependent haloes of the Milky Way and LMC. In particular, we need an improved understanding of BFEs so that we can interpolate over the properties of both haloes in a computationally efficient way. Detecting the existence of and characterizing these effects would be a spectacular test of the collisionless nature of dark matter and would offer another window to probe alternative dark matter and alternative gravity models.  

\section*{Acknowledgements}
SL would like to thank L. Whitehouse for useful discussion relating to the analysis in this paper. We would like to thank the referee for their careful and insightful review that helped improve the clarity of this manuscript.
This paper includes data obtained with the Anglo-Australian Telescope in Australia. We acknowledge the traditional owners of the land on which the AAT stands, the Gamilaraay people, and pay our respects to elders past and present.\\
MSP acknowledges grant support from Segal ANR-19-CE31-0017 of the French Agence Nationale de la Recherche (\url{https://secular-evolution.org}).
This work has made use of data from the European Space Agency (ESA) mission
{\it Gaia} (\url{https://www.cosmos.esa.int/gaia}), processed by the {\it Gaia}
Data Processing and Analysis Consortium (DPAC,
\url{https://www.cosmos.esa.int/web/gaia/dpac/consortium}). Funding for the DPAC
has been provided by national institutions, in particular the institutions
participating in the {\it Gaia} Multilateral Agreement.\\
For the purpose of open access, the author has applied a Creative Commons Attribution (CC BY) licence to any Author Accepted Manuscript version arising
from this submission.
\\
\textit{Facilities:} Anglo-Australian Telescope (AAOmega+2dF), Institute for Astronomy {\it cuillin} supercomputer (\url{https://cuillin.roe.ac.uk}), University of Surrey \textit{Eureka} High Performance Computing cluster\\
\textit{Software:} \textsc{astropy} \citep{Astropy...2013, Astropy...2018}, \textsc{exp} \citep{petersen..exp..21}, \textsc{ipython} \citep{ipython}, \textsc{jupyter} \citep{jupyter}, \textsc{matplotlib} \citep{matplotlib}, \textsc{mpi4py} \citep{mpi4py}, \textsc{numpy} \citep{numpy}, \textsc{pandas} \citep{pandas...paper, pandas...software}, \textsc{pybind11} \citep{pybind11}, \textsc{scipy} \citep{scipy}

\section*{Data Availability}

The data underlying this article is published in \citet{Koposov2022}. 
The stream models will be shared on reasonable request to the corresponding author. 
A {\sc python} interface to integrate orbits and access the expansion model for the simulation can be found here: \href{https://github.com/sophialilleengen/mwlmc}{https://github.com/sophialilleengen/mwlmc}.


\bibliographystyle{mnras}
\bibliography{references}



\appendix
\section{BFE reconstructions by radial order}
\label{sec:bfemoments}

\begin{figure*}
    \begin{subfigure}[c]{\textwidth}
    \centering
	\includegraphics[width=1.\textwidth]{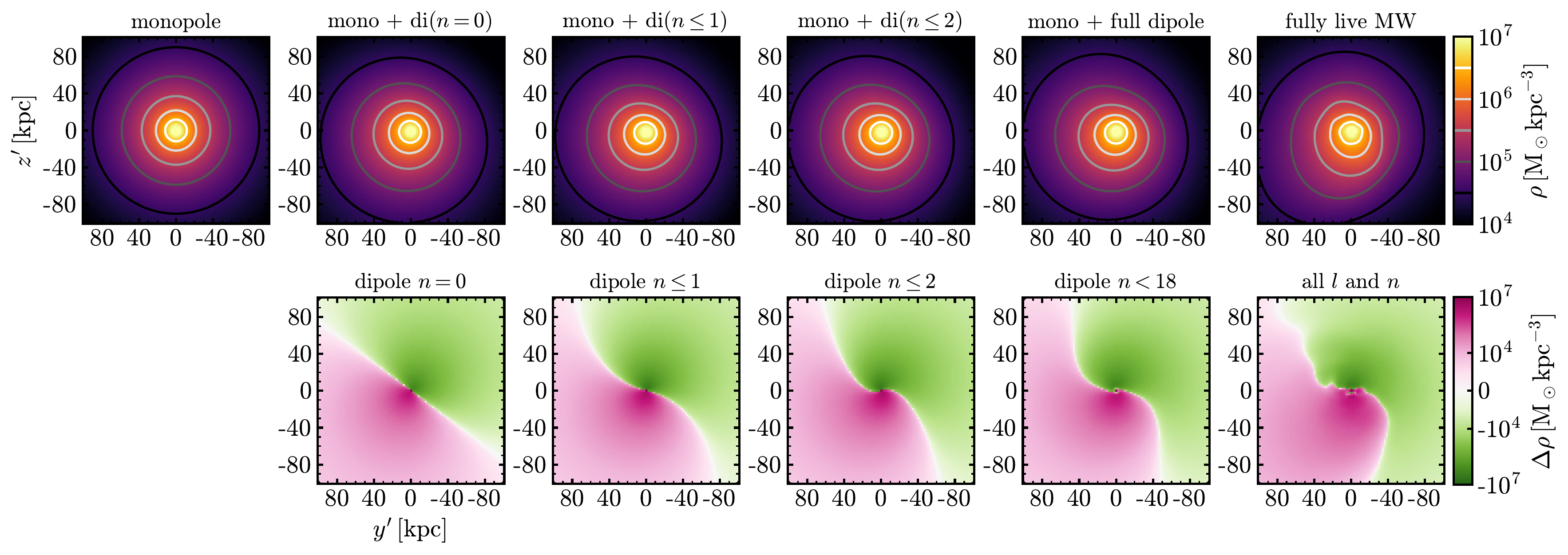}
    \end{subfigure}
    \\
    \begin{subfigure}[c]{\textwidth}
    \centering
	\includegraphics[width=1.\textwidth]{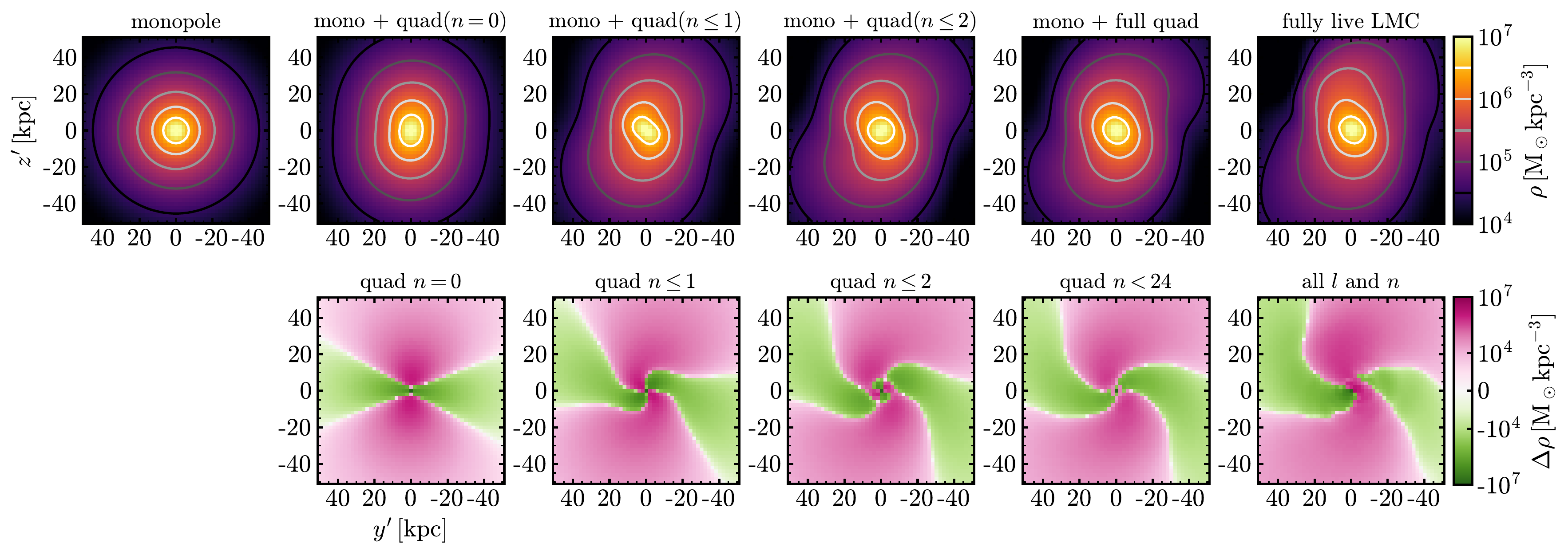}
    \end{subfigure}
    ~
    \caption{Milky Way and LMC density and change in density as a function of basis function $n$ order at the present day ($T=0$) in our model, for the largest-power harmonic subspace for each component. These are shown in the orbital plane of the LMC, defined in Appendix~\ref{app:rot_mat}. The upper two rows show the Milky Way reconstruction when adding successive \textit{dipole} radial functions The lower two rows show the LMC reconstruction when adding successive \textit{quadrupole} functions. Both the Milky Way and LMC demonstrate the convergence of the series with the addition of radial orders (compare with the full reconstructions in the right-most row).}
    \label{fig:app:ncoefs}
\end{figure*}

The spherical basis function expansions used in this paper correspond to harmonic indices $l$ and $m$ (which correspond to standard spherical harmonics), and radial index $n$, which broadly defines the spatial scale that a function is both sensitive to, and influences. The lowest-order radial function ($n=0$) is sensitive to the largest scales, the next radial function ($n=1$) is sensitive to slightly smaller scales, and so on. However, given that the basis is \textit{global}, one cannot directly map radial functions to `resolution'. To assist in physical interpretation, in this Appendix we visually demonstrate the effects on reconstructing the density field when including and excluding radial orders. We restrict our detailed analysis with radial order to the largest-power harmonic subspaces for each component, as discussed in Sections~\ref{sec:streams:tracks} and ~\ref{sec:streams_forces} (i.e. dipole for the Milky Way, quadrupole for the LMC). 

Figure~\ref{fig:app:ncoefs} shows the density reconstructions for the Milky Way and LMC, as well as the change in density for isolated $n$ orders. Beginning with the upper row, we show the reconstruction of the Milky Way when including successively more terms, from the monopole (including all radial orders) in the left-most panel, to the full density reconstruction over all functions (harmonic and radial) in the right-most panel. 
Intermediate panels (left to right) add dipole radial functions $n=0$, $n\le1$, $n\le2$, and all radial dipole orders (second from right). 
One sees that the addition of functions adds features to the overall density profile that deform the Milky Way away from the initially spherical shape. 
To further illustrate the role different functions play in determining the overall structure, in the second row of panels, we show the contribution of specific sets of functions. 
The functions are shown in parallel with the total density reconstructions in the upper row, such that the left-most panel in the second row shows the contribution of the dipole $n=0$ term, and the right-most panel shows the contribution of all non-monopole harmonic ($0<l\le6$) and radial orders ($n<18$).

We then show the same analysis for the LMC in the bottom two rows of panels of Figure~\ref{fig:app:ncoefs}, except we focus on the quadrupole rather than the dipole. 
Moving from left to right in the panels reconstructing the total density, one sees that the addition of radial terms acts to twist the inner isodensity contours. 
The ability of multiple radial orders to act together is even more apparent in the density contribution of the isolated non-monopole functions (the bottom row of Figure~\ref{fig:app:ncoefs}), where the full position angle of the LMC dipole requires the first few terms ($n\le2$) before it resembles the full reconstruction (the right-most panel).
Figure~\ref{fig:app:ncoefs} also shows that the large-scale density fields of both the Milky Way and LMC are well described with the $n\leq2$ expansions, while the higher order terms are responsible for smaller-scale features.

\section{Coordinate system transformation}\label{app:rot_mat}
\begin{figure}
    \centering
    \includegraphics[width=0.8\columnwidth, trim={1cm 2cm 0.1cm 0.6cm},clip]{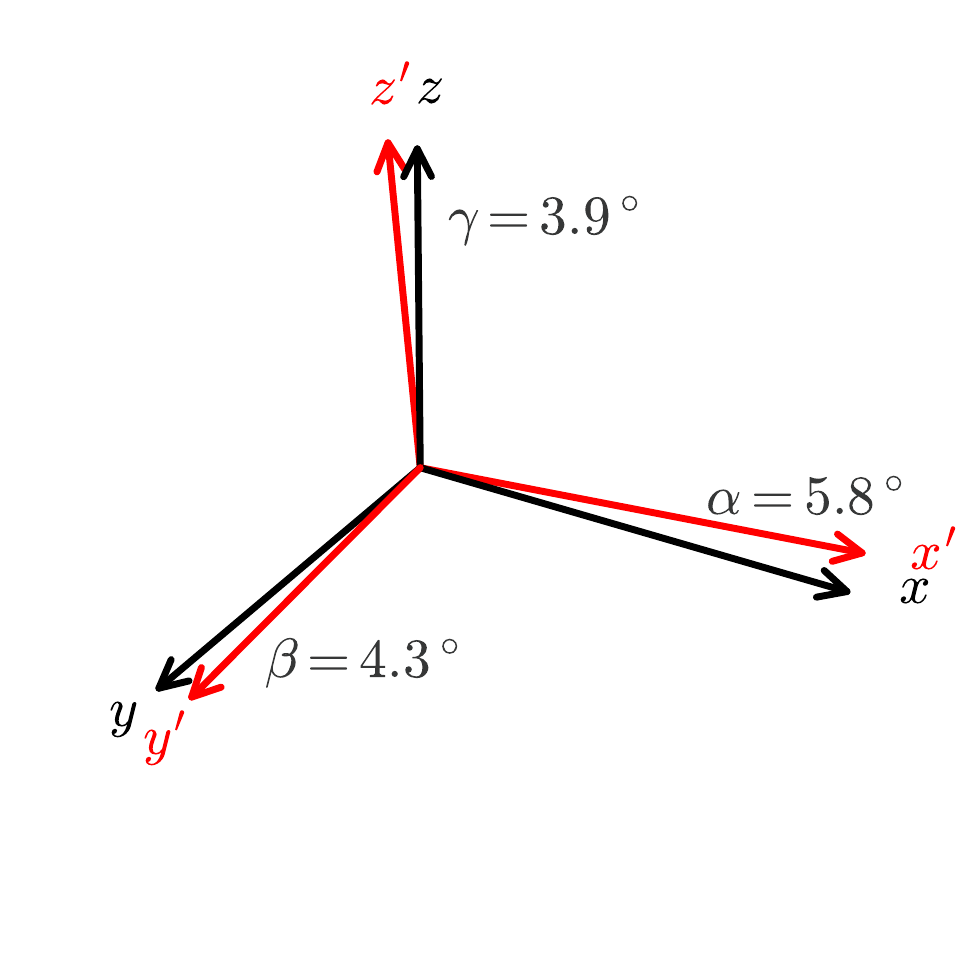}
    \caption{Comparison of coordinate systems: Galactocentric \textit{(black)} and the rotated system where the orbital plane of the LMC is aligned with the $y'z'$-plane such that the recent past LMC orbits in both frames are matching \textit{(red)}.
    The frames are close and the angles between the directions are small, but there are important features visible in the rotated frame that are not as clear in the Galactocentric frame.}
    \label{fig:app:coordinatesystems}
\end{figure}
In order to study the effects of the deformations induced by the LMC, we rotate our coordinate system so that it is aligned with the LMC's orbital plane. First, we calculate the angular momentum of the LMC 
\begin{equation}
    \vec{L}_\mathrm{LMC} = \vec{x}_\mathrm{LMC} \times \vec{v}_\mathrm{LMC} 
    = 
    \begin{pmatrix}
    -15211   \\
    1133  \\
    -1033 
    \end{pmatrix}
    {\rm ~kpc~km~s}^{-1}
\end{equation}
from its current position $\vec{x}_\mathrm{LMC}$ and current velocity $\vec{v}_\mathrm{LMC}$. This angular momentum is defined by the angles $\theta = \arctantwo(L_y, L_x)$\footnote{We choose the sequence of arguments ($y,x$) for the \texttt{arctan2} functions in line with the {\sc numpy} definition.} and $\phi = \arcsin(L_z/ |\vec{L}|)$ that are used for the rotations. The first rotation is defined by the rotation matrices
\begin{equation}
    M_1(\theta)= 
    \begin{bmatrix}
    \cos(\chi(\theta)) & -\sin(\chi(\theta)) & 0 \\
    \sin(\chi(\theta)) & \cos(\chi(\theta)) & 0 \\
    0 & 0 & 1
    \end{bmatrix}
\end{equation}

\begin{equation}
    M_2(\phi) = 
    \begin{bmatrix}
    \cos(\phi) & 0 & -\sin(\phi) \\
    0 & 1 & 0 \\
    \sin(\phi) & 0 & \cos(\phi) 
    \end{bmatrix}
\end{equation}
where
$\chi(\theta) = \pi - \theta$. 
The matrix $M_{21}(\phi, \theta) = M_2(\phi)M_1(\theta)$ aligns the orbital plane of LMC with the $y'z'$-plane, pointing towards the $-x$ direction. The position of the LMC in this new frame is calculated by $\vec{x}_\mathrm{LMC, rot}(\phi, \theta) = M_{21}(\phi, \theta) \times \vec{x}_\mathrm{LMC}$.
The angle $\psi$ is the angle between the $y$ and $z$-components of the LMC's current position in Galactocentric coordinates and in the by $M_{21}$ rotated frame 
\begin{multline}
    \psi(\phi, \theta) = -\arctantwo\left(z_\mathrm{LMC, rot}(\phi, \theta), y_\mathrm{LMC, rot}(\phi, \theta)\right)\\
    + \arctantwo\left(z_\mathrm{LMC}, y_\mathrm{LMC}\right).
\end{multline}
With this angle, we rotate the frame around $x'$ so that the LMC is at the same angle in both frames and their recent past orbits roughly match using the rotation matrix
\begin{equation}
    M_3(\phi, \theta) = 
    \begin{bmatrix}
    1 & 0 & 0 \\
    0 & \cos(\psi(\phi, \theta)) & -\sin(\psi(\phi, \theta)) \\
    0 & \sin(\psi(\phi, \theta)) & \cos(\psi(\phi, \theta)) 
    \end{bmatrix}
\end{equation}
and the combined matrix 
\begin{align}
    M(\phi, \theta) &= M_3(\phi, \theta) M_{21}(\phi, \theta) \\
    &= 
    \begin{bmatrix}
    0.99496001 & -0.07410371 &  0.06755159\\
    0.0741843 &  0.99724368 & 0.00131805 \\
    -0.06746307 &  0.00369986 &  0.99771491\\
    \end{bmatrix}
    .
\end{align}
The product of $M$ and any vector (e.g. position, velocity, or force) rotates that vector into the orbital plane of the LMC. The transpose of $M$ rotates the vector back into the Galactocentric coordinate frame.
The new coordinate system is calculated by 
\begin{equation}
    \vec{r}' = M(\phi, \theta) \vec{r}.
\end{equation}
In order to show how this new coordinate system looks, we transform the unit vectors in the prime coordinates system to vectors in the Galactocentric system in Figure~\ref{fig:app:coordinatesystems}. We note that the required rotations are relatively small and the Galactocentric cartesian axes are within $6^\circ$ of the rotated Cartesian axes.

\section{Stream observables}\label{app:all_observables}
\begin{figure*}
    \begin{subfigure}[c]{0.49\textwidth}
    \centering
    \includegraphics[width=\textwidth]{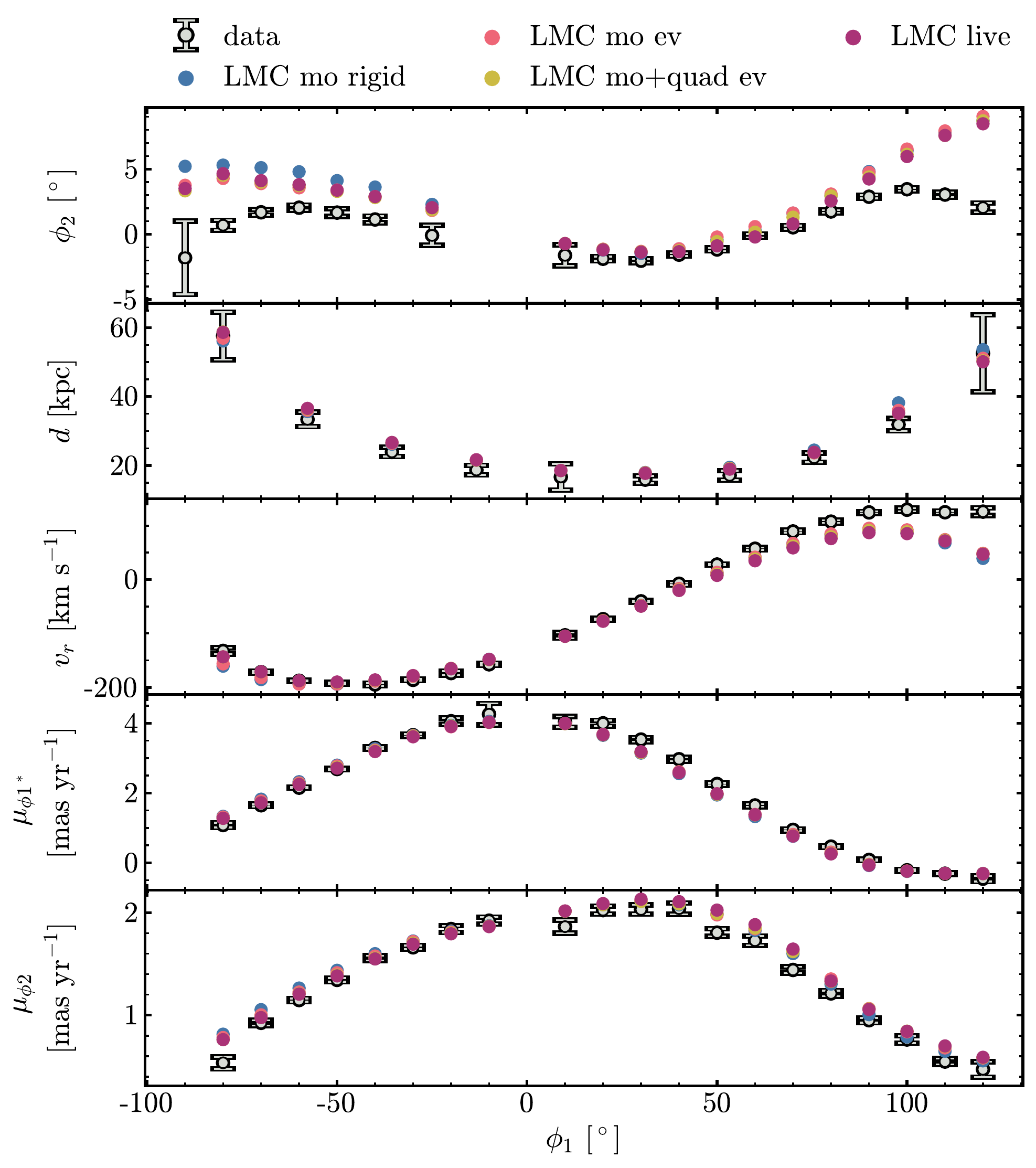}
    \end{subfigure}
    ~
    \begin{subfigure}[c]{0.49\textwidth}
    \centering
	\includegraphics[width=\textwidth]{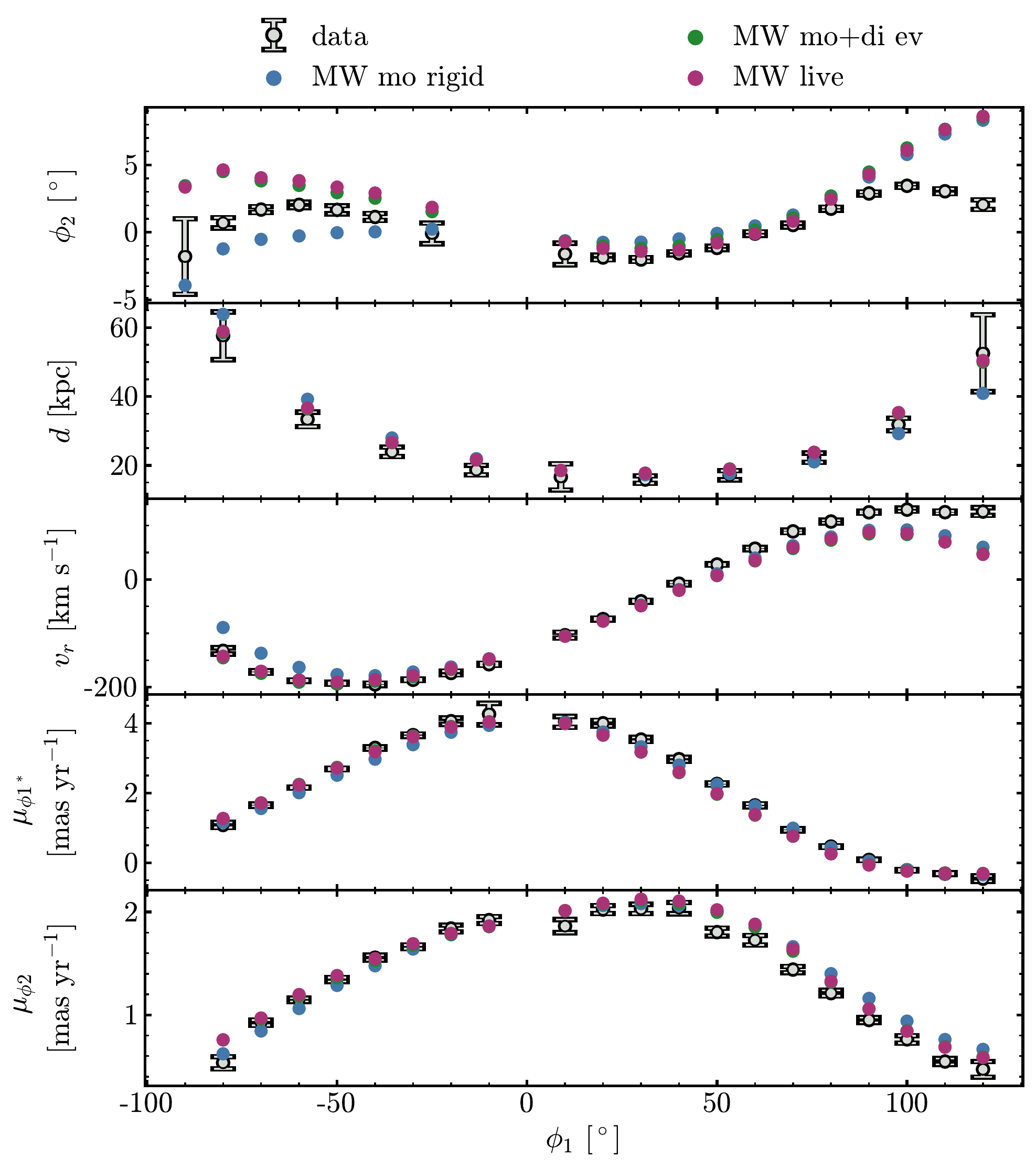}
    \end{subfigure}
    \caption{Observables of the OC stream in different LMC and Milky Way potentials. 
    The panels and data are the same as in Figure~\ref{fig:streams:full_live_stream}. 
    We show the moments with the highest impact on the stream, identified in Figures~\ref{fig:streams:track_oc} and \ref{fig:streams:f_int}. 
    \textit{Left panel:} The OC stream is evolved in a live Milky Way and different moments from the LMC are shown. 
    The moments with the highest impact are turning on the time-dependence (blue to red) and then including the quadrupole (red to yellow and purple). 
    \textit{Right panel:} The LMC is kept live while the OC stream is evolving in different Milky Way moments. 
    The most important moment is the dipole (blue to green and purple). 
    For both potential setups (Milky Way live vs LMC live), the track is the observable with the biggest changes as seen in Figure~\ref{fig:streams:track_oc}. 
    The other observables mostly do not differ significantly more than the data uncertainty.}
    \label{fig:app:streams:observables}
\end{figure*}
Figure~\ref{fig:app:streams:observables} shows the observables (i.e. stream track, distance, proper motions, and radial velocity) for the data and the modelled streams in the BFE moments of each galaxy with the largest impact on the stream. The most affected observable is the stream track (which is the focus of this work) but we see other observables are also affected and thus fitting all observables should provide stringent constraints on these deforming haloes.


\bsp	
\label{lastpage}
\end{document}